\documentclass[aps,prb,twocolumn,superscriptaddress,floatfix]{revtex4-2}

\usepackage{amsmath, amssymb, amsfonts}
\usepackage{bbold}
\usepackage{hyperref}
\hypersetup{colorlinks=true,
		    linkcolor=blue,
		    citecolor=blue,
		    allcolors=blue}
\usepackage{color}
\usepackage{graphics, graphicx}

\usepackage{dcolumn}
\newcolumntype{d}[1]{D{.}{.}{#1}}
\usepackage{multirow}
\usepackage[percent]{overpic}
\usepackage{wasysym}
\usepackage{comment}

\DeclareMathOperator{\Tr}{Tr}
\makeatletter
\g@addto@macro\bfseries{\boldmath}
\makeatother

\usepackage{bm, bbm, braket, mathtools, comment}
\usepackage{tcolorbox}
\tcbuselibrary{theorems}
\usepackage{orcidlink}
\usepackage{babel}
\usepackage{pifont}

\usepackage{pgfplots}

\newcommand{\AG}{A$\Gamma$}
\newcommand{\KG}{K$\Gamma$}
\newcommand{\AK}{A$K$}
\newcommand{\FK}{F$K$}
\newcommand{\RHOA}{$\rho_A$}
\newcommand{\RHOB}{$\rho_B$}
\newcommand{\SPTA}{SPT$_\alpha$}
\newcommand{\SPTB}{SPT$_\beta$}

\newcommand{\FMU}{FM$_{U_6}$}
\newcommand{\RSU}{RS$_{U_6}$}

\newcommand{\HU}{$H_{\mathrm{K}\Gamma}^{U_6}$}
\newcommand{\HKG}{$H_{\mathrm{K}\Gamma}$}

\usepackage{pdfpages} 
\usepackage{pgffor} 

\makeatletter
\AtBeginDocument{\let\LS@rot\@undefined}
\makeatother

\def\supplementfilename{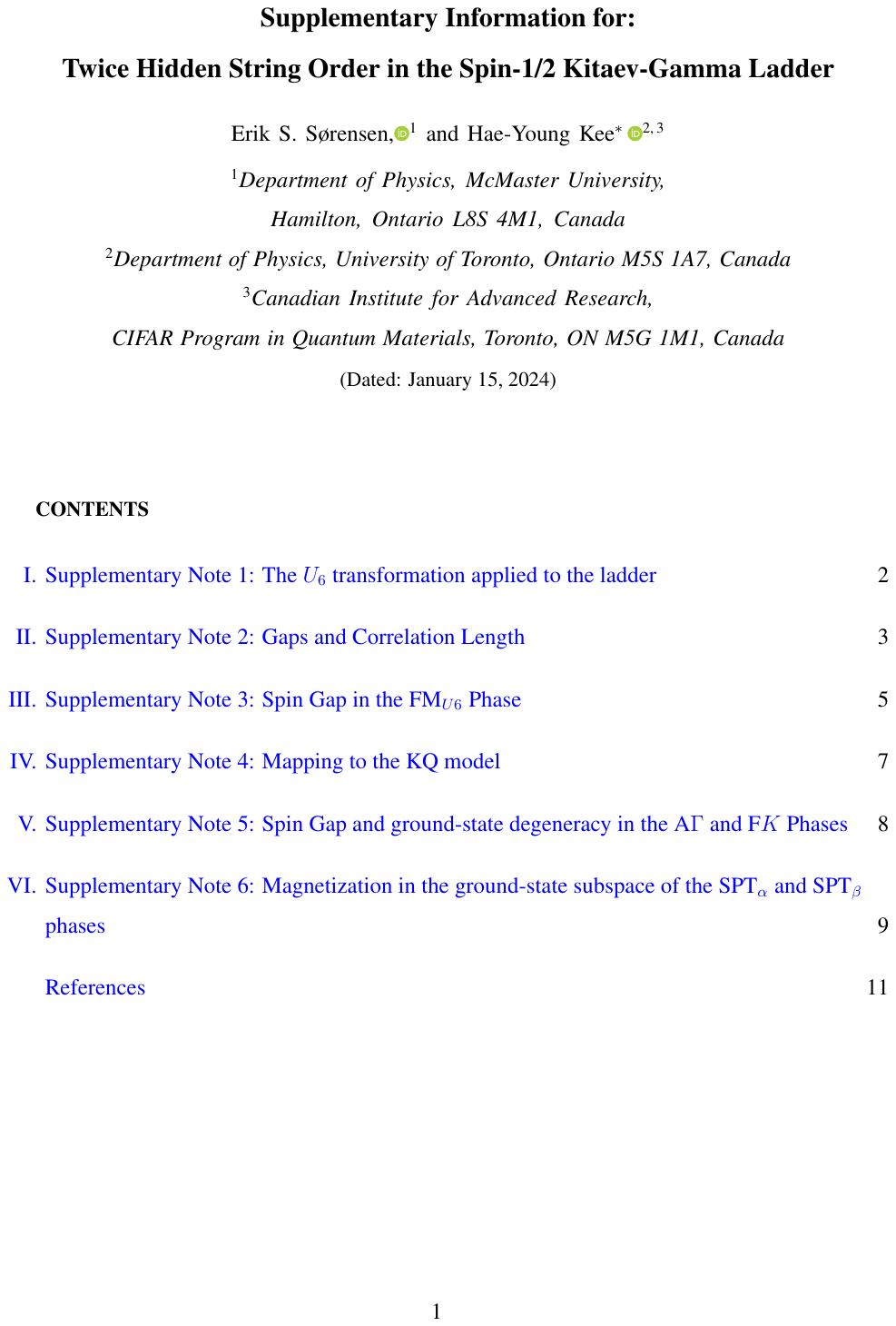}

\pdfximage{\supplementfilename}
\def\numbersupplementpages{\the\pdflastximagepages}

\newif\ifarXiv
\arXivtrue

\begin{document}

\title{Twice Hidden String Order and Competing Phases in the Spin-1/2 Kitaev-Gamma Ladder}
\author{Erik S. S{\o}rensen\,\orcidlink{0000-0002-5956-1190}}
\affiliation{Department of Physics and Astronomy, McMaster University, Hamilton, Ontario L8S 4M1, Canada}
\author{Hae-Young Kee$^*$\,\orcidlink{0000-0002-8248-4461}}
\affiliation{Department of Physics, University of Toronto, Ontario M5S 1A7, Canada}
\affiliation{Canadian Institute for Advanced Research, CIFAR Program in Quantum Materials, Toronto, ON M5G 1M1, Canada}
\date{\today}

\begin{abstract}
Finding the Kitaev spin liquid in candidate materials involves understanding the entire phase diagram, including other allowed interactions. One of these interactions, called the Gamma ($\Gamma$) interaction, causes magnetic frustration and its interplay with the Kitaev ($K$) interaction is crucial to comprehend Kitaev materials. Due to the complexity of the combined \KG\ model, quasi-one dimensional models  have been investigated. While several disordered phases are found in the 2-leg ladder, the nature of the phases are yet to be determined. Here we focus on the disordered phase near the antiferromagnetic $\Gamma$ limit (denoted by \AG\ phase) next to the ferromagnetic Kitaev phase. We report a distinct non-local string order parameter characterizing the \AG\ phase, different from the string order parameter in the Kitaev phase. This string order parameter becomes evident only after two unitary transformation, referred to as a twice hidden string order parameter. The related entanglement spectrum, edge states, magnetic field responses, and the symmetry protecting the phase are presented, and its relevance to the two-dimensional Kitaev materials is discussed. Two newly identified disordered phases in the phase diagram of  \KG\ ladder is also reported.

\end{abstract}
\maketitle

\section*{Introduction}

Over the course of investigating spin $S$=1/2 two-dimensional (2D) honeycomb Kitaev materials~\cite{balents2014review,rau2016review,Takagi2019review,trebst2022review,Rouso2023ARCMP}, candidates of long-sought quantum spin liquids, an additional bond-dependent spin exchange term named Gamma ($\Gamma$) interaction~\cite{rau2014prl} was found along with the bond-dependent Kitaev ($K$) interaction~\cite{kitaev2006anyons,jk2009prl}.
Unlike the standard Heisenberg interaction, both the Kitaev and the $\Gamma$ interactions are highly nontrivial and extremely frustrated. While the same sign of $K$ and $\Gamma$ cancels the frustration leading to magnetically ordered phases, the regions with different signs of these interactions are highly frustrated.\cite{Rouso2023ARCMP}
Since these two interactions are known to dominate over other symmetry-allowed interactions in the emerging candidate material $\alpha$-RuCl$_3$, 
the combined Kitaev-Gamma (\KG) model has attracted considerable attention in the theoretical community and it is widely accepted~\cite{janssen2017model,winter2017review,hermanns2018review} that a realistic description of $\alpha$-RuCl$_3$ should be sought in the regime with antiferromagnetic (AFM) Gamma ($\Gamma\mathord{>}$0) and ferromagnetic (FM) Kitaev ($K\mathord{<}$0) interactions. Understanding the phases that arise in this frustrated regime of the 2D honeycomb \KG-model is therefore of crucial importance.

This particular region of the \KG-model has been extensively studied by a range of numerical methods~\cite{Rouso2023ARCMP}.
Despite detailed studies, the nature of the phase next to the FM Kitaev spin liquid 
arising due to AFM $\Gamma$ interaction  
remains controversial. 
Most 
studies have found that it is in a disordered phase~\cite{Catuneanu2018npj,Gohlke2018prb,Gohlke2020PRR,Yamada2020prb,Lee2020Magnetic,
Luo2021npj,Gordon2019,Yilmaz2022PRR}, 
denoted by K$\Gamma$SL for \KG\ spin liquid, or nematic paramagnets,  but functional renormalization approaches found magnetically ordered phases~\cite{Buessen2021} while
variational Monte Carlo calculations~\cite{Wang2019prl} observed a narrow disordered phase next to the FM Kitaev spin liquid with most of the antiferromagnetic $\Gamma$ region dominated by a zig-zag ordered phase.

Motivated by such discrepancy, 
another approach to investigate the 2D limit of the \KG\ model was taken  by starting from low-dimensional models
with the hope of furthering the understanding of the honeycomb model by determining the phases of $n$-leg (brick-wall) models. Despite the obvious
challenge in connecting the two limits, it is reasonable to expect that 
potential spin liquid phases arising in the honeycomb model should correspond to regions where the $n$-leg models display disordered phases.
Such an approach was employed earlier for the pure Kitaev model~\cite{feng2007characterization}. Disordered phases in the anisotropic Kitaev 1-leg chain were found, and they were characterized by non-local string order parameters (SOPs)\cite{feng2007characterization}.  It has also been shown that the isotropic Kitaev 2-leg ladder model exhibits a disordered phase, characterized by an unconventional SOP different from that of the anisotropic chain Kitaev phase~\cite{Catuneanu2018ladder}.

The one-dimensional (1D) chain and ladder version of the \KG\ model were investigated numerically with very high precision,  as the reduced dimensionality allows accessing bigger system sizes~\cite{Wang2020a,Wang2020b,luo2021prb,SorensenPRR2023a,SorensenPRR2023b}.
In the 1D chain model,  it was found that the pure Gamma model belongs to a Luttinger liquid phase governed by the gapless hidden SU(2) Heisenberg chain, a fact revealed after a 6-site transformation, i.e, a duality mapping~\cite{Chaloupka2015hidden,Wang2020a}.
A study of the same \KG\ model on a quasi 1D ladder using DMRG and iDMRG techniques\cite{Gordon2019,sorensen2021prx}
found a magnetically disordered phase, possessing a small gap near the AFM pure Gamma limit. This phase surrounding the pure AFM Gamma point next to the FM Kitaev phase (denoted by \FK) was referred to as the
\AG\ phase. 
Even though the \AG\ phase in the ladder occurs in the same part of the phase diagram as the proposed $K\Gamma$SL in the 2D limit, a
distinct name  was introduced, as it is yet to be determined how the \AG\ phase is connected to the 2D limit $K\Gamma$SL.

While it is clear that there is no magnetic order in this phase, the precise nature of the \AG\ phase has not yet been settled due to its complex nature. 
The presence of a gap indicates that the \AG-phase is likely a symmetry protected topological (SPT) phase~\cite{Wen1989,Gu2009,WenRMP2017}. 
If so, it is of importance to identify a corresponding SOP, edge states, and the symmetry that protects this phase, which are characteristic of the SPT, and
to determine how this phase respond to an external magnetic field. 
If strong evidence for a non-trivial SPT nature of the \AG\ phase can be established, it would
establish a next step to 
the proposed $K\Gamma$SL in the 2D limit. In the following sections, we will systematically examine these inquiries and provide comprehensive responses. 
To perform a thorough analysis, we start by reviewing the full phase diagram prior to focusing on the nature of the \AG\ phase. As shown below, we also report two other disordered phases.

\section*{Results}

\subsection*{Model}\label{sec:model}

\begin{figure}
  \includegraphics[width=\columnwidth,clip]{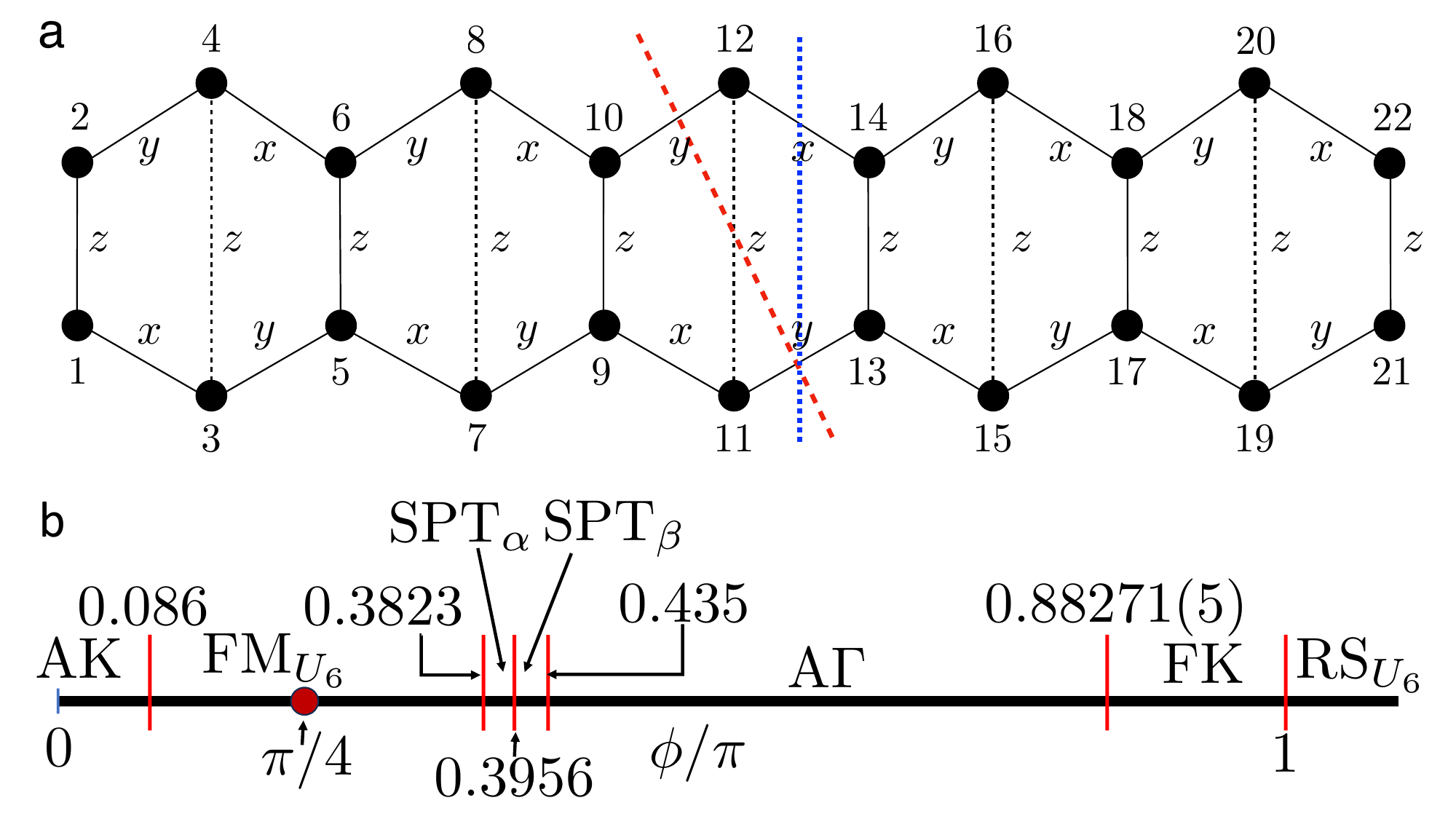}
   \caption{{\bf The \KG\ ladder and corresponding phase diagram.}
   {\bf a}
   A strip of the \KG\ honeycomb lattice corresponding to a two-leg \KG\ ladder with alternating $x$ and $y$ bonds along the leg and $z$-bond between the chains, with the numbering of the sites used throughout the paper.
   The dotted $z$-bonds arise from imposing periodic boundary conditions in
   the direction perpendicular to the ladder. However, all $z$-bonds are taken to have equal strength. 
   The dashed red line indicates the partition used for \RHOB\ while the dashed blue line indicates the partition used for \RHOA.
   {\bf b} Schematic phase diagram of the \KG\ ladder, Eq.~(\ref{eq:HKG}), for $\Gamma\mathord{>}$0. 
 }
  \label{fig:ladder}
\end{figure}
The \KG\ Hamiltonian is given by
\begin{equation}
   H_{K\Gamma} = \sum_{\langle i,j\rangle_{\gamma \in (x,y,z)}}  K S_i^\gamma S_j^\gamma + \Gamma\left(S^\alpha_iS^\beta_j + S^\beta_iS^\alpha_j\right) 
   \label{eq:HKG}
\end{equation} 
where $(\alpha, \beta)$ takes on the values $(y,z)/(x,z)/(x,y)$ for $\gamma = x/y/z$, and $\langle i, j \rangle$ refers to the nearest neighbor sites.
An alternative representation of the honeycomb lattice is as a brick-wall lattice~\cite{feng2007characterization}, and its two-leg limit with periodic boundary conditions is simply a ladder shown in Fig.~\ref{fig:ladder}{\bf a}.
The dotted bonds indicate Kitaev $z$-bonds arising from
periodic boundary conditions, and we shall always take such bonds to be identical to the regular (solid) $z$-bonds, in which case the honeycomb strip
can be viewed as a regular rectangular ladder.

We  parameterize the model by taking
\begin{equation}
K=\cos\phi, \ \ \ \mathrm{and}\ \ \ \  \Gamma\mathord{=}\sin\phi,
\end{equation}
and interpolate between the Kitaev and $\Gamma$ interactions by varying $\phi$.
Our main interest is in the region with $\phi/\pi\in [0,\pi]$ where $\Gamma\mathord{>}$0 and
the Kitaev term, $K$, changes from AFM to FM at $\phi$=$\pi$/2, as this region is relevant to most two-dimensional (2D) Kitaev candidate materials.
The total number of sites in the ladder (including both legs) is denoted by $N$.

\subsection*{Phase Diagram}\label{sec:zerofield}

A full phase diagram is shown in Fig.~\ref{fig:ladder}{\bf b}. It is obtained by various quantities presented in the next subsection.
Moving from $\phi\mathord{=}0$ to $\pi$, the AFM Kitaev phase (denoted by AK), 
a FM phase denoted by FM$_{U_6}$, \AG, and \FK\ phases are found consistent with the earlier works\cite{Gordon2019,sorensen2021prx}. 
At the special point $\phi=\pi/4$, the \FMU\ phase can be mapped to the ferromagnetic Heisenberg ladder by applying a local
unitary $U_6$ transformation (see Supplementary Note 1), and is therefore gapless.
However, for $\phi\mathord{\neq}\pi/4$ a small gap appear, as we show in  Supplementary Note 3. 
Surprisingly, two additional phases denoted by \SPTA\ and \SPTB\ can be identified between the \FMU\ and \AG\ phases.
As we will show below, they are magnetically disordered and display the characteristics of SPT phases,
i.e., doubled entanglement spectrum.
Beyond $\phi=\pi$, the rung singlet phase denoted by \RSU\ phase delineates the \FK\ phase. 

In addition to the expected Kitaev phases \AK\ and \FK\, the appearance of the \FMU\ and \RSU\ phases are well established in the \KG\ honeycomb and n-leg models.  After the local $U_6$ transformation~\cite{Chaloupka2015hidden}, corresponding to local spin rotations, at $\phi\mathord{=}\pi/4$ and $5\pi/4$, i.e., $K$=$\Gamma$, the \KG\ 2D honeycomb model is equivalent to the FM and AFM Heisenberg model, respectively. 
The application of the $U_6$ transformation is specified in Supplementary Note 1.
To understand the nature of the other three phases, \AG, \SPTA\ and \SPTB, 
we first performed a detailed analysis of the entanglement spectrum.

\begin{figure}[t]
        \includegraphics[width=\columnwidth]{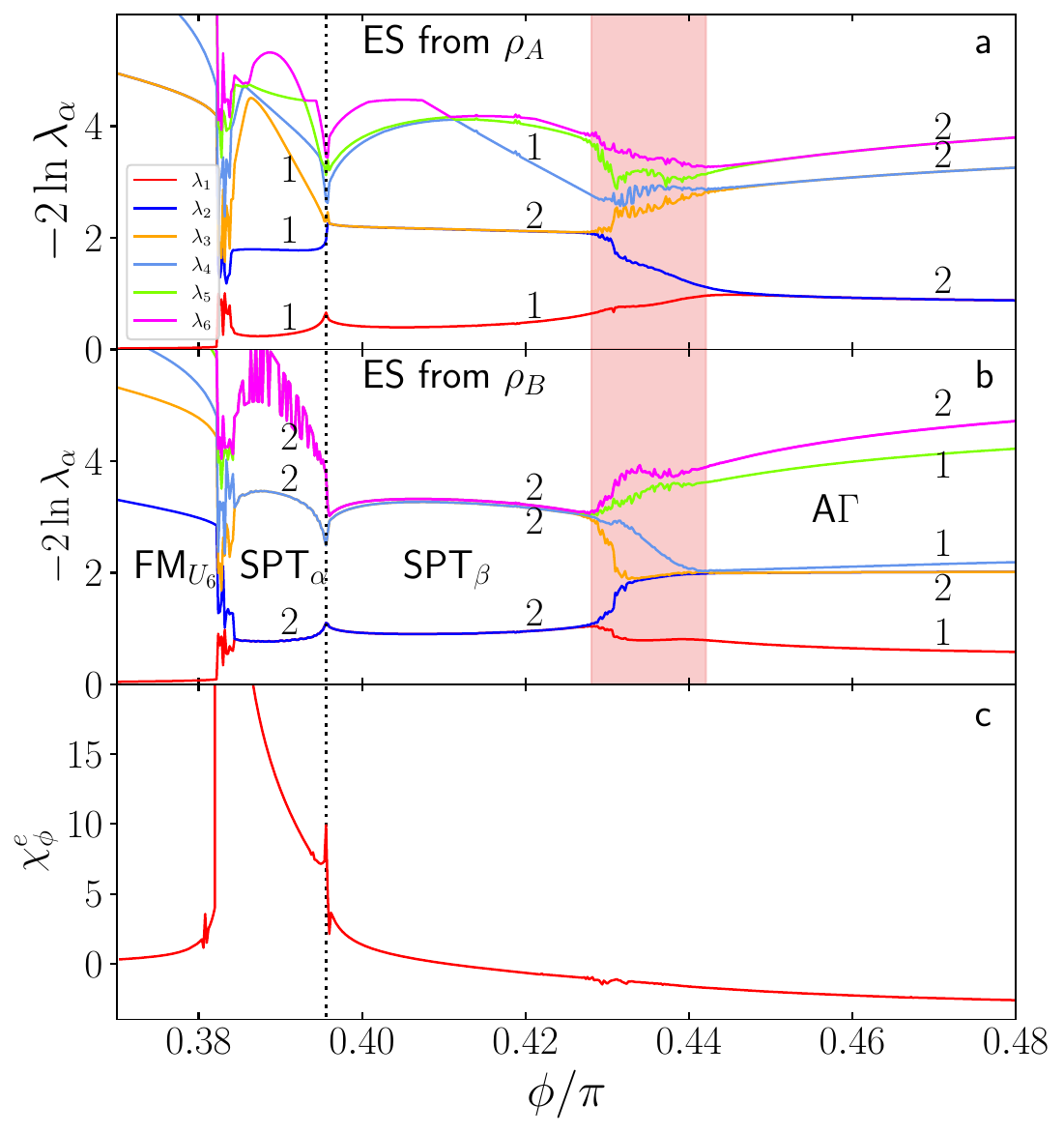}
  \caption{{\bf Entanglement spectrum and susceptibility for $\phi/\pi\mathord{<}0.5$}.
  iDMRG results for: {\bf a} Entanglement spectrum from \RHOA, red dashed line in Fig.~\ref{fig:ladder}{\bf a}. {\bf b} Entanglement spectrum from \RHOB,  blue dashed line in Fig.~\ref{fig:ladder}{\bf a}. 
  The numbers refer to the degeneracy of the eigenvalue.  {\bf c} $\chi^e_\phi$. 
  Two distinct phases, \SPTA\ and \SPTB, are visible between the \FMU\ and \AG\ phase. The red shading between $\phi/\pi$=0.428-0.442 denotes a transitional region of limited convergence due to a field instability.
  }
  \label{fig:ShortEigs}
\end{figure}

\begin{figure}[t]
        \includegraphics[width=\columnwidth]{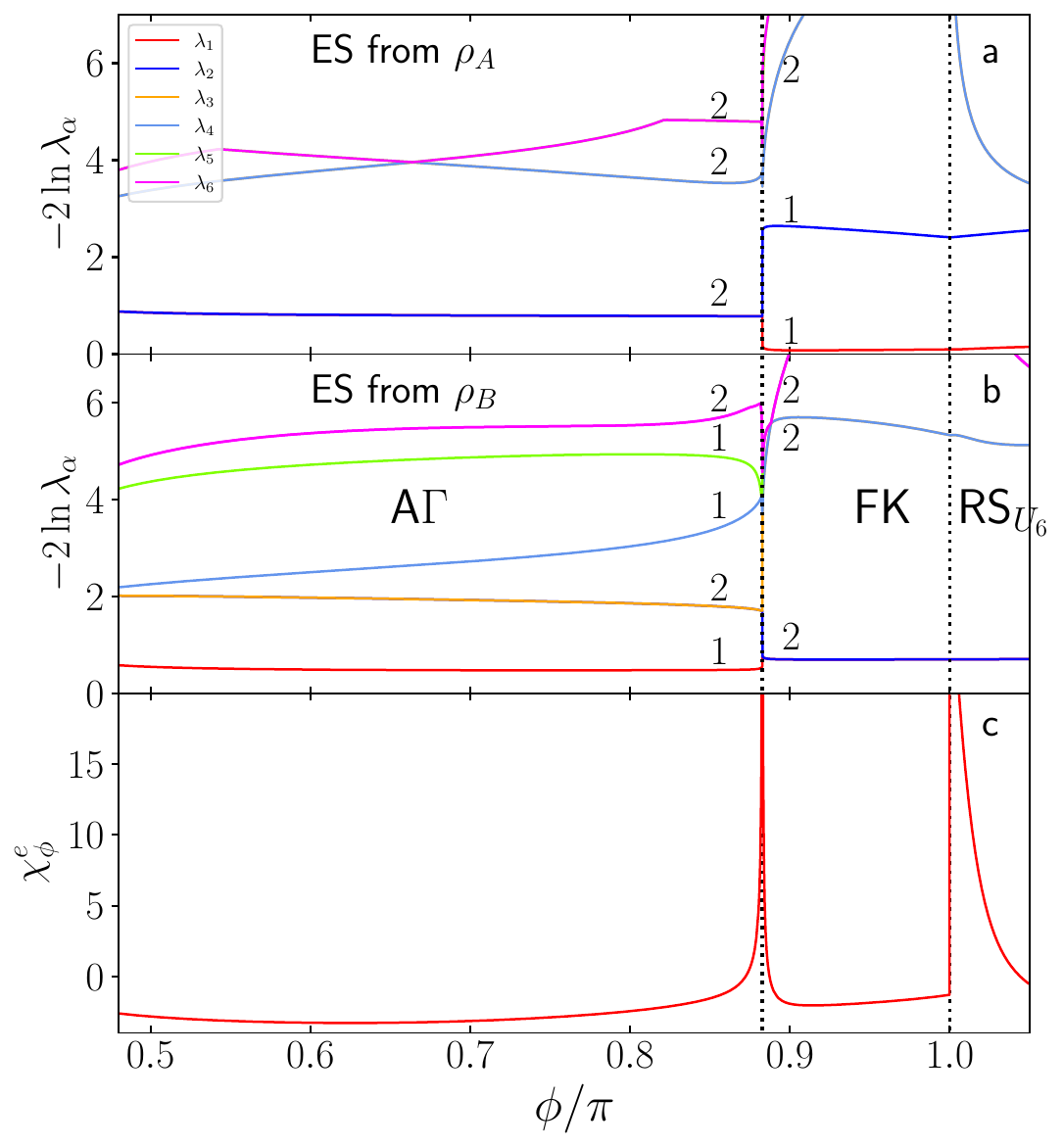}
  \caption{{\bf Entanglement spectrum and susceptibility for $\phi/\pi\mathord{>}0.5$}.
  iDMRG results for: {\bf a} Entanglement spectrum from \RHOA, red dashed line in Fig.~\ref{fig:ladder}{\bf a}. {\bf b} Entanglement spectrum from \RHOB,  blue dashed line in Fig.~\ref{fig:ladder}{\bf a}. The numbers refer to the degeneracy of the eigenvalue. {\bf c} $\chi^e_\phi$. The \AG\, \FK\ and \RSU\ phases are clearly delineated.
  }
  \label{fig:FullEigs}
\end{figure}

\subsection*{Entanglement Spectrum}
Our results for the entanglement spectrum, as well as for the susceptibility, $\chi^e_\phi$ are shown in Figs.~\ref{fig:ShortEigs}
and \ref{fig:FullEigs}. Due to the complexity of the phase diagram, we split it into two regions centered around the \AG\ phase: Fig. ~\ref{fig:ShortEigs} represents the left part of the \AG\ and Fig. \ref{fig:FullEigs} the right part of the \AG\ phase.
We consider two different partitions shown in Fig.~\ref{fig:ladder}{\bf a} corresponding to \RHOA\ 
and \RHOB. In both Fig.~\ref{fig:ShortEigs} and Fig.~\ref{fig:FullEigs} iDMRG results are shown for both partitions with
\RHOA\ shown in panel (a) and \RHOB\ in  panel (b) with $\chi^e_\phi$ in panel (c).

To the left of the \AG-phase, 
two other phases \SPTA\ and \SPTB\ 
are clearly separated from the FM$_{U_6}$ and \AG, as noted from
their entanglement spectrum for \RHOA. 
Furthermore, the double degeneracy  of the entanglement spectrum shown in Fig. \ref{fig:ShortEigs} (b) from \RHOB\ is a clear signal
of a SPT phase~\cite{Pollmann2010,Pollmann2012a,Schuch2011,Chen2011,WenRMP2017}.
Note that there is only a weak signature in $\chi^e_\phi$ of the transition between the two phases, visible at $\phi$=$0.3956\pi$ in Fig.~\ref{fig:ShortEigs}(c), corresponding to a small discontinuity in $\chi^e_\phi$.
While the FM$_{U_6}$ phase appears abruptly at $\phi$=$0.3823\pi$, the transitions between the \AG\ phase and \SPTB\ phase at $\phi\sim 0.435\pi$ is 
not discernible in $\chi^e_\phi$.
The blurriness of the transition is likely due to a field induced phase that pinches off to a single point at zero field
at the A$\Gamma$-\SPTB\ transition, thereby obscuring it. We note that, the quantum critical points (QCPs) are immediately noticeable in the entanglement spectra.

Moving to the right within the \AG\ phase which encompass the point $\Gamma$ =1, 
the transition to the \FK\ phase from the \AG\ phase 
occurs at $\phi\mathord{=}0.88271(5)\pi$, which is clearly visible in the entanglement spectra as well as $\chi^e_\phi$ as shown in 
in Fig.~\ref{fig:FullEigs}. The transition between the RS$_{U_6}$ and FK phases at $\phi$=$\pi$  is less apparent in the structure of the entanglement spectrum in Fig.~\ref{fig:FullEigs} (a) and (b).
However,  from the results for $\chi^e_\phi$ in  Fig.~\ref{fig:FullEigs}(c) 
this transition is immediately visible, and it was also noted
that the second derivative of $\lambda_1$ clearly detects the transition~\cite{sorensen2021prx}.

Similar to the \AK\ phase, none of the phases \SPTA, \SPTB, \AG\ and \FK\ has any long-range magnetic ordering, nor is there any indication
of nematic (quadropolar) or chiral ordering. As discussed in Supplementary Note 2, all four phases are gapped with a finite sizeable correlation length. The difference between
them is captured in the entanglement spectrum.
In the \SPTA, \SPTB\ and \FK\ (and \AK) phases, the entanglement spectrum
have all entries doubled when considering \RHOB. 
For the \AG-phase, the same applies to the spectrum for \RHOA. 
Since an entanglement spectrum where all eigenvalues have degeneracy larger than one is a signature of a topological non-trivial phase, in the next section, we investigate the projective symmetry analysis to confirm their non-trivial topology prior to presenting associated non-local SOPs.

\subsection*{Projective Symmetry Analysis}\label{sec:proj}
With the \SPTA, \SPTB, \AG\ and \FK\ phases as potential SPT phases, it is of considerable interest to investigate the projective representations~\cite{Chen2011,Liu2011,Chen2011b,Liu2011b,Liu2012,Chen2013,WenRMP2017}
$U$, of a site symmetry $\mathcal{R}$. This has been done for $S$=1 chains~\cite{Chen2011,Liu2011,Chen2011b,Liu2011b,Liu2012,Chen2013}
and ladders~\cite{Chen2015} and also for $S$=1/2 ladders~\cite{Liu2012,Ueda2014,Kariyado2015,Ogino2021,Ogino2022}.
The matrices $U$ can be obtained from the generalized transfer matrices in iDMRG calculations as described in Ref.~\onlinecite{Pollmann2012b}.

\begin{figure}[t]
  \includegraphics[width=0.9\columnwidth]{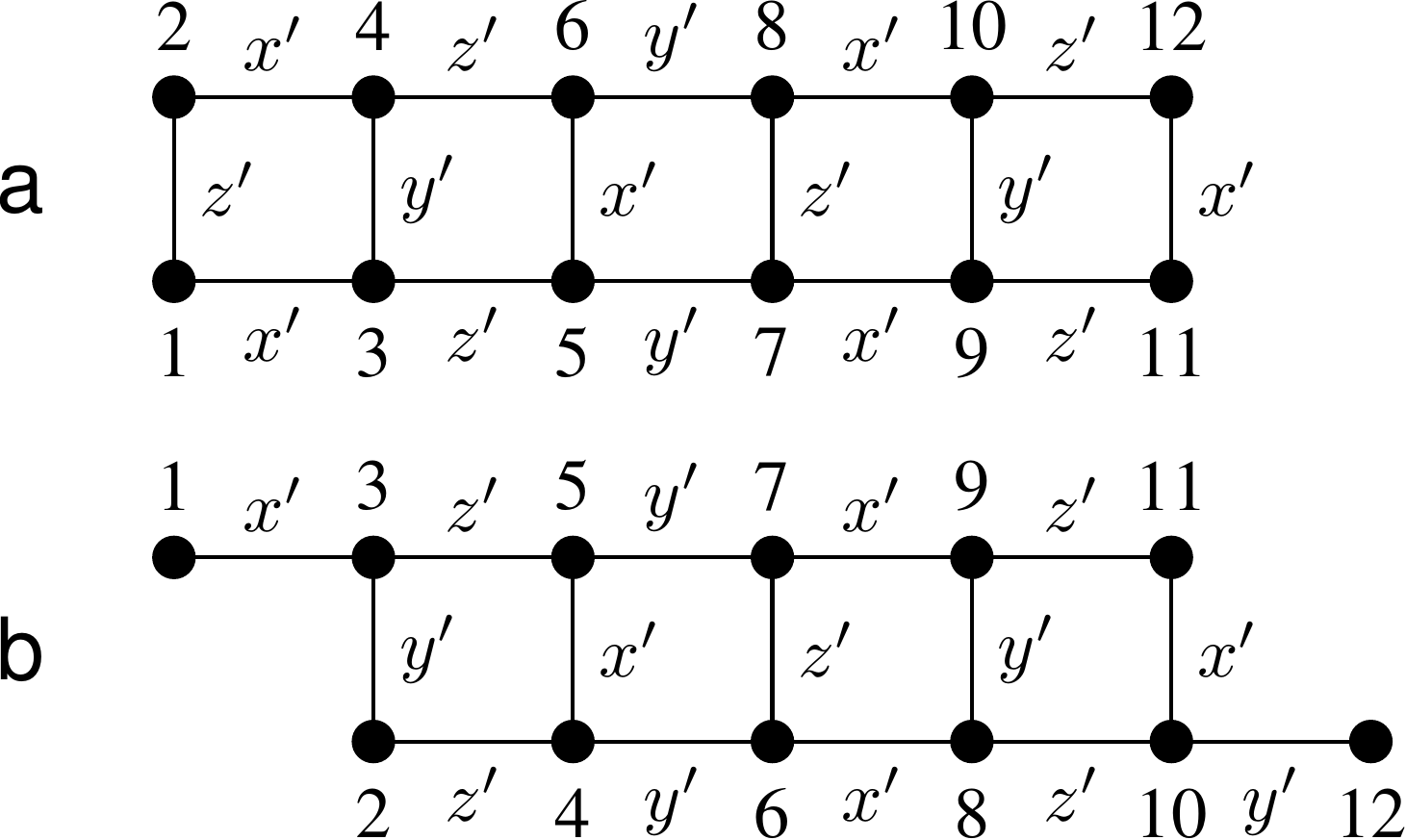}
  \caption{{\bf Unit cells of the $U_6$ transformed ladder.} Two unit cells of \HU, the \KG\ ladder after the $U_6$ local transformation with $N\mathord{=}6n$ sites.
  {\bf a} Two regular unit cells with regular open boundary conditions (OBC). 
  {\bf b} Two slanted unit cells with slanted OBC.}
   \label{fig:kgusix}
\end{figure}

For the \KG\ ladder, it is a significant simplification to consider the transformed model obtained after applying the $U_6$ transformation.
This transformation maps the original Hamiltonian \HKG\ to the transformed ladder, denoted by \HU.
Under the $U_6$ transformation the $x$, $y$ and $z$-bonds of the \KG-ladder are transformed into anisotropic Heisenberg bonds, $x'$, $y'$ and $z'$ in the following manner:
\begin{eqnarray}
  x'\ :\ -KS^x_iS^x_j-\Gamma(S^y_iS^y_j+S^z_iS^z_j)\nonumber\\
  y'\ :\ -KS^y_iS^y_j-\Gamma(S^x_iS^x_j+S^z_iS^z_j)\nonumber\\
  z'\ :\ -KS^z_iS^z_j-\Gamma(S^x_iS^x_j+S^y_iS^y_j)\label{eq:u6}
\end{eqnarray}
In the transformed ladder \HU\ model, 
the definition of $x'$, $y'$ and $z'$-bonds, can pictorially be represented
as shown in Fig.~\ref{fig:kgusix}(a).
See Supplementary Note 1.
We will consider two different open boundary conditions (OBC) as shown in Fig.~\ref{fig:kgusix}. The unit cells depicted in panel (a) are referred to as regular unit cells with regular OBC, while the ones in panel (b) are referred to as slanted unit cells with slanted OBC.

In order to understand the symmetries of the matrix product state (MPS) wave-function, it is useful to write it in the canonical form~\cite{Vidal2003b,Vidal2007,PerezGarcia2007,Orus2008}:
\begin{equation}
    |\Psi\rangle = \sum_{j_1,\ldots,j_N}M^{[1]}_{j_1}\Lambda^{[2]}M^{[2]}_{j_2}\ldots \Lambda^{[N]}M^{[N]}_{j_N} |j_1,\ldots,j_N\rangle,
\end{equation}
where the $M^{[n]}_{j_n}$ are complex matrices and the $M^{[n]}$, real, positive, square diagonal matrices.
In the iDMRG formulation, the set of matrices on any unit cell becomes the same 
$M^{[n]}_{j}$=$M_{j}$, $M^{[n]}$ = $M$ for all $n$, although they may vary within the unit cell.
For the translationally invariant state, it can be shown~\cite{PerezGarcia2008,Pollmann2010} that for any (site) symmetry operation $g$, represented in the spin basis by the unitary matrix, $\Sigma_{jj'}(g)$, the $M_j$ matrices must transform as~\cite{Pollmann2010,Pollmann2012b}:
\begin{equation}
   \sum_{j'} \Sigma_{jj'}(g)M_{j'} = e^{i\theta}U^\dagger(g) M_j U(g), 
\end{equation}
where the unitary matrix $U(g)$ commutes with the $M$ matrices and $e^{i\theta}$ is a phase factor. With $D$ denoting the bond dimension, the $U$ matrices form a $D$-dimensional projective representation of the symmetry group of the wave-function, and they can be determined from the unique eigenvector of the generalized transfer matrix~\cite{Pollmann2010,Pollmann2012b} with eigenvalue $|\lambda|$=1,
where the  generalized transfer matrix is defined as 
\begin{equation}
T^\Sigma_{\alpha\alpha';\beta\beta'}=\sum_j\left(\sum_{j'}\Sigma_{jj'}M_{j',\alpha\beta}\right)(M_{j,\alpha'\beta'})^*\Lambda_{\beta}\Lambda_{\beta'}
\end{equation}
If the largest eigenvalue is $|\lambda|<$1, the symmetry is not a property of the state being considered. The projective representation is reflected in the fact that if
$\Sigma(g)\Sigma(h)$=$\Sigma(gh)$, then 
\begin{equation}
   U(g)U(h)=e^{i\phi(g,h)}U(gh),
\end{equation}
where the phases $\phi(g,h)$ are characteristic of the topological phase.

Let us now consider the site symmetries, 
$\mathcal{R}^x$ and $\mathcal{R}^y$ defined as:
\begin{equation}
    \mathcal{R}^x_l=e^{i\pi S^x_l}\ ,\ \ \ \mathcal{R}^y_l=e^{i\pi S^y_l}\ ,\ \ \ \mathcal{R}^z_l=e^{i\pi S^z_l}.\label{eq:R}
\end{equation}
The Hamiltonian \HU\ is invariant under the operators $\prod_l\mathcal{R}^\gamma_l$, with $\gamma\mathord{=}x,y,z$ with distinct
quantum numbers for the low-lying states. 
If these symmetries are respected, their representations can differ by a phase, $\phi(\mathcal{R}_x,\mathcal{R}_y)$ that must obey
$e^{i\phi(\mathcal{R}_x,\mathcal{R}_y)}$=$\pm 1$:
\begin{equation}
U(\mathcal{R}^x)U(\mathcal{R}^y)=\pm U(\mathcal{R}^y)U(\mathcal{R}^x).
\end{equation}
Furthermore, the non-trivial value $e^{i\phi(\mathcal{R}_x,\mathcal{R}_y)}$=-1 can only occur if all eigenvalues of the entanglement spectrum are at least twice degenerate~\cite{Pollmann2010}. 
The phase factor can then be isolated by defining~\cite{Pollmann2012b}:
\begin{equation}
\mathcal{O}_\mathrm{Z_2\times Z_2}\equiv\frac{1}{D}\Tr\left( U(\mathcal{R}^x)U(\mathcal{R}^y) U^\dagger (\mathcal{R}^x)U^\dagger (\mathcal{R}^y)\right),\label{eq:OZ2}
\end{equation}
with $D$ the bond dimension, with similar definitions for other pairs of operators in Eq.~(\ref{eq:R}). 
In the above definitions it is understood that the transformations are applied throughout the lattice and in order to obtain the matrices $U$, 
generalized transfer matrices representing the relevant unit cell has to be considered. 

The site symmetries, $\mathcal{R}^x$, $\mathcal{R}^y$ and $\mathcal{R}^z$, forming the dihedral group, $D_2$, are respected by both
of the unit cells in Fig.~\ref{fig:kgusix}(a) and \ref{fig:kgusix}(b). 
Using generalized transfer matrices obtained from unit cells of the shape shown in Fig~\ref{fig:kgusix}(a) when studying the \AG-phase and of the slanted shape shown in Fig.~\ref{fig:kgusix}(b) when studying the \FK, \SPTA, and \SPTB\ phases, we obtain
\begin{equation}
\mathcal{O}_\mathrm{Z_2\times Z_2} = -1,
\end{equation}
for the \SPTA, \SPTB, \AG, and \FK\ phases.

Similar analysis can be made for the 
time-reversal $(\mathrm{TR})$ symmetry, defined by 
$M_j\to \sum_{j'}\left[e^{i\pi S^y}\right]_{jj'}M^{\star}_{j'}$, 
with $\star$ denoting complex conjugation. In this case, it can be established that~\cite{Pollmann2010} $U_\mathrm{TR}U_\mathrm{TR}^\star$=$e^{i\phi(\mathrm{TR},\mathrm{TR})}\mathbb{1}$ where the phase $\phi(\mathrm{TR},\mathrm{TR})$ cannot be absorbed into the definition of $U_{\mathrm{TR}}$. One should note that for most other symmetries,
with the notable exception of inversion, similar considerations will lead to $U^2$=$e^{i\phi}\mathbb{1}$ in which case the phase 
$\phi$
in fact can be absorbed into the definition of $U$. 
For instance, this is the case for $U\left(\mathcal{R}^\alpha\right)$ discussed above. However, for time reversal, the phase factor $e^{i\phi(\mathrm{TR},\mathrm{TR})}$ can directly be extracted by defining
\cite{Pollmann2012b}:
\begin{equation}
\mathcal{O}_\mathrm{TR}\equiv\frac{1}{D}\Tr\left( U_\mathrm{TR}U_\mathrm{TR}^\star \right),\label{eq:OTR}
\end{equation}
and again one finds that $\phi(\mathrm{TR},\mathrm{TR})$=0,$\pi$, so that  $\mathcal{O}_\mathrm{TR}$=$\pm$1.
As an example, for the $S$=1 Heisenberg spin chain in the Haldane phase it is known that $\mathcal{O}_\mathrm{Z_2\times Z_2}$=$-1$, $\mathcal{O}_\mathrm{TR}$=$-1$~\cite{Pollmann2010,Pollmann2012a}. 

Using generalized transfer matrices obtained from unit cells of the shape shown in Fig~\ref{fig:kgusix}(a) for the \AG-phase and of the slanted shape for the \FK, \SPTA, and \SPTB\ phases, we obtain
\begin{equation}
\mathcal{O}_\mathrm{TR}=-1,
\end{equation}
consistent with the presence of a doubled entanglement spectrum in all cases. 
As was the case for $\mathcal{O}_\mathrm{Z_2\times Z_2}$ , if the unit cells are interchanged, one finds instead $\mathcal{O}_\mathrm{TR}$=1. 

\begin{table}[h!]
  \caption{{\bf Summary of projective analysis}. The superscript $\mathrm{regular}$ refers to the unit cell from Fig.~\ref{fig:kgusix}{\bf a}, while the superscript $\mathrm{slanted}$ refers to the unit cell from Fig.~\ref{fig:kgusix}{\bf b}. 
  Negative values indicate that the state transforms non-trivially.}
    \label{tab:table1}
    \begin{ruledtabular}
    \begin{tabular}{l|c|c|c|c|}
      Phase & $\mathcal{O}_\mathrm{TR}^\mathrm{regular}$ & 
      $\mathcal{O}_\mathrm{Z_2\times Z_2}^\mathrm{regular}$ & 
      $\mathcal{O}_\mathrm{TR}^\mathrm{slanted}$&
      $\mathcal{O}_\mathrm{Z_2\times Z_2}^\mathrm{slanted}$\\ 
      \hline
      \SPTA\   & 1 & 1  &  -1& -1\\ 
      \SPTB\  &  1 & 1  &  -1& -1\\ 
      \AG\    & -1 &  -1  &  1& 1\\ 
      \FK\     & 1 &  1  &  -1& -1\\ 
    \end{tabular}
    \end{ruledtabular}
\end{table} 
A summary of our results from the projective analysis are provided in Table~\ref{tab:table1}, negative values indicate that the state transforms non-trivially. For all 4 phases, it is seen that a unit cell can be chosen for which the state transforms non-trivially under both  the $\mathrm{TR}$ and $\mathcal{O}_\mathrm{Z_2\times Z_2}$ symmetries.

Based on the above analysis of entanglement spectrum and projective symmetry, we conclude that \AG\ phase is an SPT phase. It is then important to further identify its SOP
that differentiates this phase from the other disordered phases.

\subsection*{Twice Hidden String Order}\label{sec:SOP}

To establish a non-local string order parameter (SOP) characterizing the \AG\ phase, we need to exploit a non-local unitary transformation that maps the original Hamiltonian with OBC to a new Hamiltonian that exhibits a local long-range order~\cite{Kennedy1992a,Kennedy1992b,Oshikawa1992}.  
We found that it is difficult to identify such non-local transformation starting from the original \HKG, but
it can be achieved by first applying the $U_6$ transformation to arrive at \HU.
It is then possible to define a non-local unitary transformation $W$, mapping \HU\ to a new local Hamiltonian.
We denote the resulting Hamiltonian, where four-spin terms appear, 
by $H_{\mathrm{KQ}}$.
For the parameters relevant for the \AG-phase,  $H_{\mathrm{KQ}}$ exhibits long-range order in the spin-spin correlation functions, corresponding to
a local order parameter. Due to the application of two separate unitary transformations, one might consider the resulting order to be twice hidden.

\begin{figure}
        \includegraphics[width=\columnwidth]{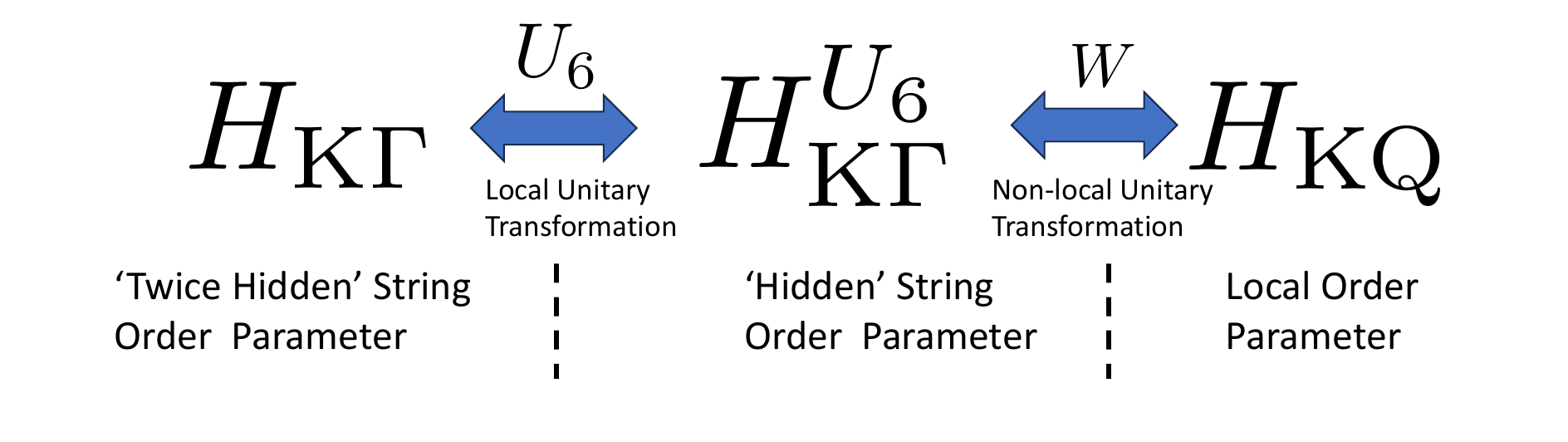}
  \caption{{\bf Pictorial view of KG transformations.}
  The two transformations from the original \KG\ model to \HU, and the subsequent transformation to $H_{\mathrm{KQ}}$ are sketched.  
  The type of order parameter for each Hamiltonian is indicated.
  }
  \label{fig:transformations}
\end{figure}

The non-local unitary operator $W$ for a  $N$-site ladder with OBC that maps \HU\ to  $H_{\mathrm{KQ}}$ takes the following form.
\begin{equation}
 W = \prod_{\substack{j+1< k \\ j\ \mathrm{odd},\  k\ \mathrm{odd} \\ j=1,\ldots N-3 \\ k=3,\ldots N-1}} w(j,k).
 \label{eq:W}
\end{equation}
With the individual $w(j,k)$ given as follows:
\begin{equation}
  w(j,k)  = e^{i\pi(S^y_j+S^y_{j+1})\cdot(S^z_k+S^z_{k+1})},
  \label{eq:w}
\end{equation}
and $W^\dagger$=$W$. 
The OBC are here crucial for the mapping to be exact. 
Evidently, all $w(j,k)$ are unitary and therefore also $W$, and  
$  \left[w(j,k),w(l,m)\right]  = 0 \ \ \forall \ j,k,l,m$.
Note that this is a different labelling of 
the 
unitary operator introduced in Ref.~\cite{Catuneanu2018ladder,sorensen2021prx}.
It can also be shown that other combinations of the spin operators $S^\alpha$ appearing in Eq.~(\ref{eq:w}) lead
to equivalent unitary operators, for instance, 
$e^{i\pi(S^y_j+S^y_{j+1})\cdot(S^x_k+S^x_{k+1})}$ 
is another valid choice.
However, the specific choice made in Eq.~(\ref{eq:w}) will influence the type of ordering that is observed in $H_{\mathrm{KQ}}$, as well as the specific form of $H_{\mathrm{KQ}}$.
Schematically, the transformations can be viewed as shown in Fig.~\ref{fig:transformations}.
The detailed form of $H_{\mathrm{KQ}}$ after the $W$ transformation on \HU\ is presented in Supplementary Note 4.

\begin{figure}[b]
  \centering
        \includegraphics[width=\columnwidth]{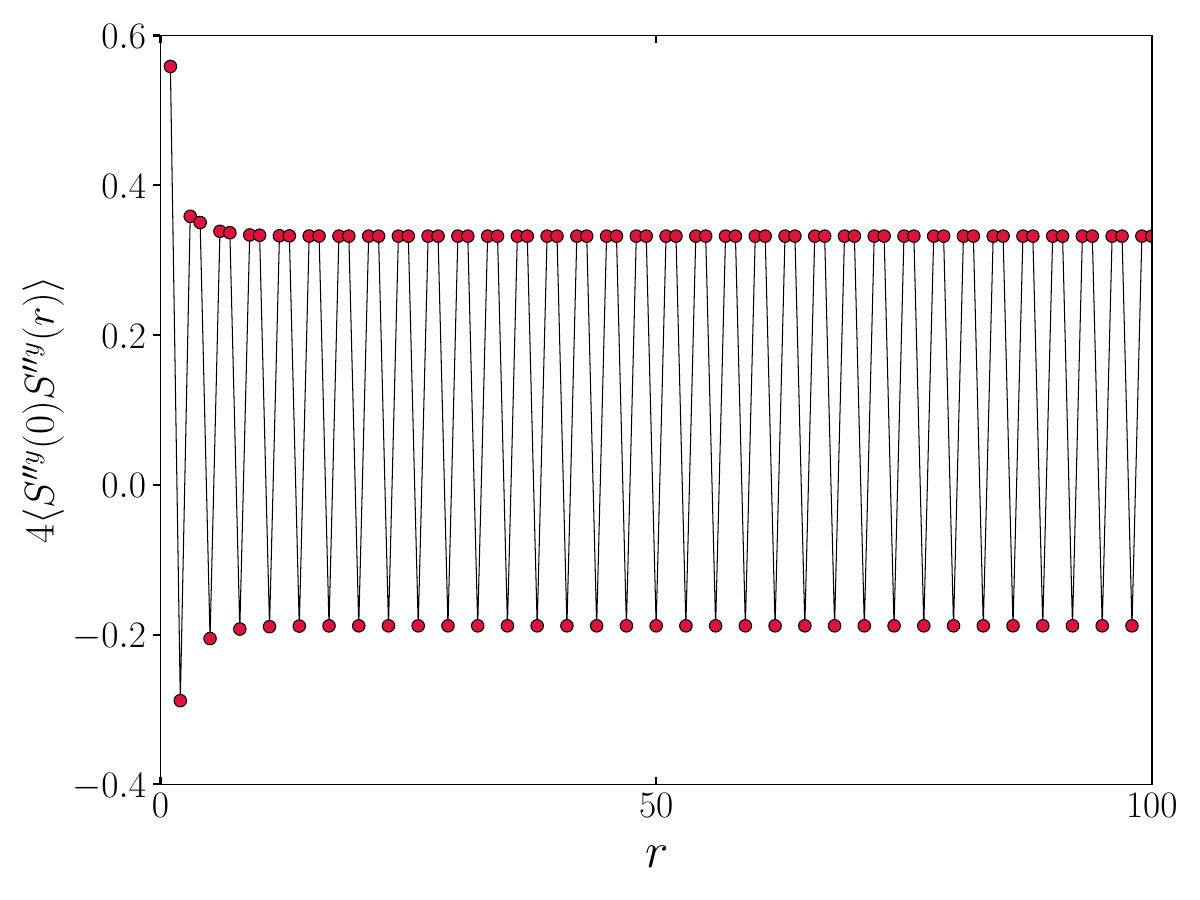}
  \caption{{\bf Spin correlations in the $H_{\mathrm{KQ}}$ model at $\phi\mathord{=}0.85\pi$.} DMRG results with $N$=384 for the correlation function $4\langle S^{\prime\prime y}(0) S^{\prime\prime y}(r)\rangle$ versus distance, $r$ along one leg of the ladder. $r$=0 corresponds to site 47.
  Results are for $H_{\mathrm{KQ}}$ with $\phi\mathord{=}0.85\pi$ and $r$ is measured in lattice spacings along the leg.
  }
  \label{fig:HKQCorr}
\end{figure}

We first discuss the ordering in $H_{\mathrm{KQ}}$ where we denote the spins by $S^{\prime\prime\alpha}$,
where the double prime represents the two transformations of spin from the original Hamiltonian.
It is then convenient to study correlation functions
of the form $4\langle S^{\prime\prime\alpha}(0)S^{\prime\prime\alpha}(r)\rangle$ along the legs of the ladder with $r$ measured in lattice spacings along the leg. To avoid boundary effects, $r$=0 is usually taken to correspond to a site in the bulk of the chain. In Fig.~\ref{fig:HKQCorr} we show results
for $4\langle S^{\prime\prime y}(0)S^{\prime\prime y}(r)\rangle$ starting from site 47 with $\phi\mathord{=}0.85\pi$. Long-range order is clearly present. 
Similar results can be obtained for the other leg of the ladder as well as for $4\langle S^{\prime\prime z}(0)S^{\prime\prime z}(r)\rangle$.
However, due to the choice of spin operators in the definition of $w$ in Eq.~(\ref{eq:w}), there is  no ordering in $4\langle S^{\prime\prime x}(0)S^{\prime\prime x}(r)\rangle$.


Using the inverse of the non-local unitary operator $W$ from Eq.~(\ref{eq:W}) the above results for  $4\langle S^{\prime\prime y}(0)S^{\prime\prime y}(r)\rangle$
is reproduced as a non-local string order correlation function in \HU\ where we denote the spin variables by $S^{\prime\alpha}$=$\sigma^{\prime\alpha}/2$. 
We find $\langle S_1^{\prime\prime y} S_{r+1}^{\prime\prime y}\rangle$ is given by
\begin{eqnarray}
  {\cal O}^y(r) &=& 4 \langle S_1^{\prime\prime y} S_{r+1}^{\prime\prime y}\rangle =  
  (-1)^{\left \lfloor{r/2}\right \rfloor }
  \nonumber\\
  &\times&\left\{ \begin{array}{rl}
  \langle\sigma_2^{\prime y}\left( \prod_{k=3}^{r}\sigma_k^{\prime y} \right)\sigma_{r+1}^{\prime y}\rangle &\mbox{ $r$ even}\\
    \\
  \langle\sigma_2^{\prime y}\left( \prod_{k=3}^{r-1}\sigma_k^{\prime y} \right)\sigma_{r+1}^{\prime y}\rangle &\mbox{ $r$ odd}
       \end{array} \right.
       \label{eq:oy}
\end{eqnarray}
Note that, to fully reproduce the results for $H_{\mathrm{KQ}}$ shown in Fig.~\ref{fig:HKQCorr} with the string correlation function in
(\ref{eq:oy}) for \HU, a relabelling of the sites needs to be done that we have skipped for clarity.
\begin{figure}[t]
        \includegraphics[width=\columnwidth]{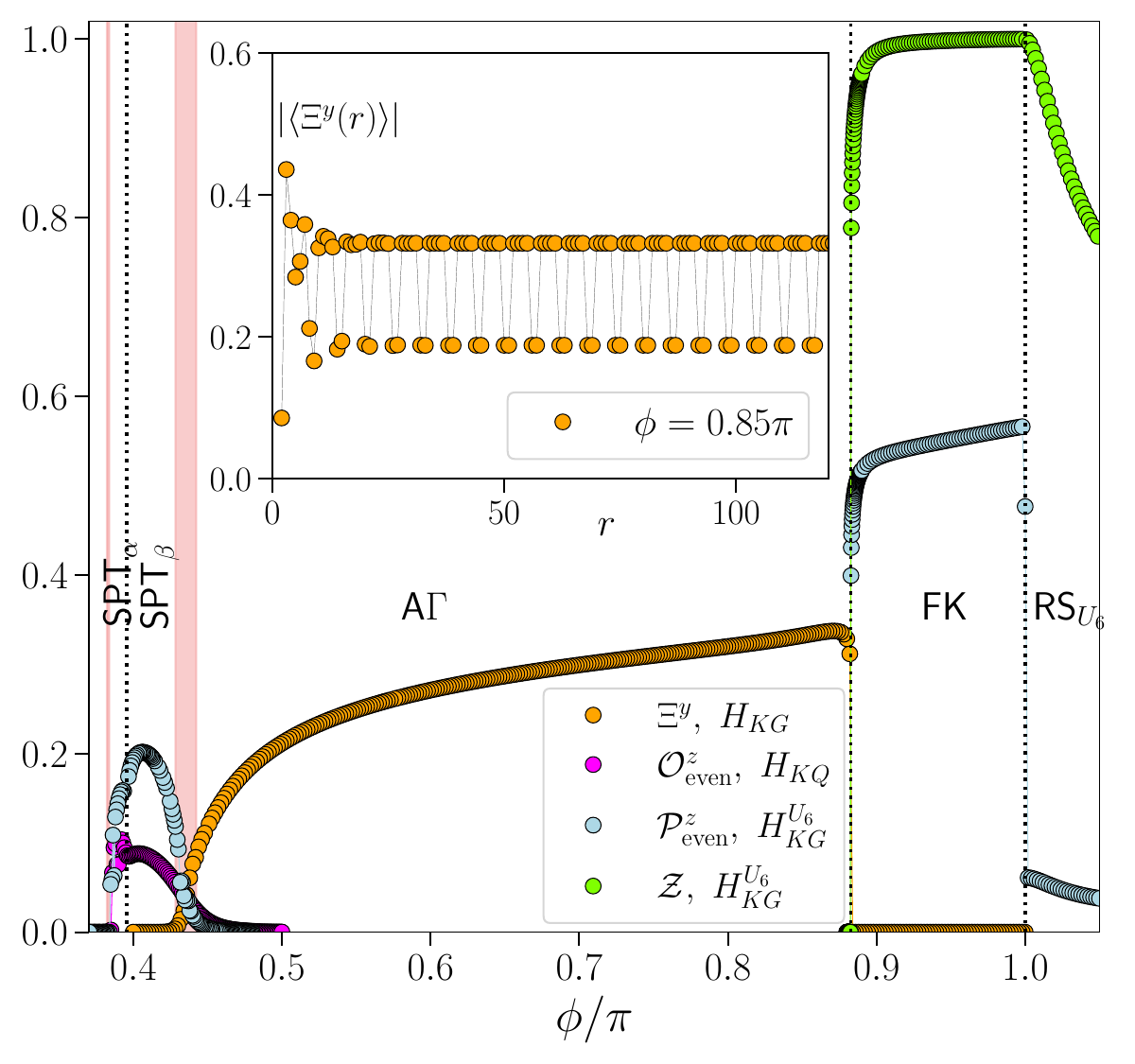}
  \caption{{\bf String order parameters in the \AG\ and surrounding phases.} iDMRG results for the string order parameters in the \HKG\ and  
  \HU\ models, shown alongside DMRG results for $H_{\mathrm{KQ}}$. Orange circles, $\Xi^y$ for \HKG. The inset shows
  $\Xi^y(r)$ versus $r$ at $\phi\mathord{=}0.85\pi$ for \HKG. Magenta circles, DMRG results for ${\cal O}_\mathrm{even}^z$ for $H_{\mathrm{KQ}}$.
  Light blue circles, ${\cal P}^z_\mathrm{even}$ for \HU. Green circles, ${\cal Z}$ for \HU.
  }
  \label{fig:AGY}
\end{figure}

We can now apply the inverse $U_6$ transformation (see Supplementary Note 1) to the above expressions for  ${\cal O}^y (r)$
to determine the string order correlation functions that define the \AG-phase in the original \HKG:
\begin{equation}
    \Xi^y(r)=U_6^{-1}({\cal O}^y(r)),
\end{equation}
with the $U_6^{-1}$ transformation detailed in Supplementary Note 1.
The SOP in \HKG\ Hamiltonian is then given by
\begin{equation}
\Xi^y = \max_{r\to\infty}\Xi^y(r).
\end{equation}
It is interesting to note that for a small $r$, for example $r=7$ $\Xi^y (r = 7)$ corresponds to the plaquette operator 
found in the pure Gamma model in the honeycomb lattice.\cite{perkins2017classical}

In Fig.~\ref{fig:AGY} we show iDMRG results for $\Xi^y$ (orange circles). The inset shows iDMRG results for $\Xi^y(r)$ versus $r$
at $\phi\mathord{=}0.85\pi$ which at large $r$ can be compared to the results for $H_{\mathrm{KQ}}$ shown in Fig.~\ref{fig:HKQCorr}. Note that the results
in Fig.~\ref{fig:HKQCorr} show the correlations along a leg, whereas the inset in Fig.~\ref{fig:AGY} show results from both legs without including
the sign and the relabelling of the sites. Due to the different methods used, there is not an exact equivalence for very small values of $r$.

We emphasize that 
the appearance of a non-local SOP in the \AG\ phase of \HKG\ is equivalent to the presence of long-range order in $H_{\mathrm{KQ}}$. Hence, $\Xi^y$ is non-zero throughout the \AG-phase and goes to zero at the critical points delineating this phase. It is absent in the other disordered phases, \SPTA, \SPTB, and \FK, and thus uniquely defines the \AG\ phase.

With the identification of the \AG-phase with regular long-range ordering in the $H_{\mathrm{KQ}}$ model, it is natural to ask if a regular (local) order parameter can also be identified for the \SPTA-\ and
\SPTB-phases in the $H_{\mathrm{KQ}}$. However, all local order parameters that we have investigated have not shown any ordering in the \SPTA-\ and \SPTB-phases for $H_{\mathrm{KQ}}$. 
Extending the string-order correlation function defined for the $S\mathord{=}1$ Haldane chain, a heuristic
string-order correlation function has been proposed for $S$=1/2 ladders by pairing two $S\mathord{=}1/2$~\cite{Kim2000,Fath2001}.
Following the numbering of Fig.~\ref{fig:ladder}{\bf a}, if $\tau_i^\alpha\mathord{=}S^\alpha_{2i}+S^\alpha_{2i+1}$ are the
sum of two diagonally situated spins, one defines~\cite{Kim2000,Fath2001}: 
\begin{equation}
  {\cal O}_\mathrm{even}^\alpha(r) = \langle \tau_i^\alpha\exp(i\pi\sum_{l=i+1}^{i+r-1}\tau_l^\alpha)\tau_{i+r}^\alpha\rangle.
  \label{eq:oeven}
\end{equation} 
The associated SOP is non-zero in the phase surrounding $\phi\mathord{=}0$ in $H_{\mathrm{KQ}}$~\cite{sorensen2021prx}, corresponding to the AFM Kitaev (\AK) phase in \HKG. The magenta points in Fig.~\ref{fig:AGY} show our results for ${\cal O}_\mathrm{even}^z$ for
the $H_{\mathrm{KQ}}$ model, which clearly is non-zero in the \SPTA-\ and \SPTB-phases. This is consistent with the nonexistence of a local order parameter in these two phases for $H_{\mathrm{KQ}}$. We note that, due to the heuristic nature of ${\cal O}_\mathrm{even}^z$, it is not clear how to associate it with a local order in a related model.
Since \HKG\ and \HU\ are related by a local unitary transformation, any ordering in the \SPTA-\ and \SPTB-phases in either model would immediately be apparent in both, and we have not observed any for either model.

Building on the above results for ${\cal O}_\mathrm{even}^z$ for $H_{\mathrm{KQ}}$ we propose a closely related heuristic string order correlation function for \HU\ in the following way: Define $l_i^\alpha\mathord{=}S^\alpha_{2i}-S^\alpha_{2i+1}$ as the difference of two diagonally
situated spins, following the numbering from Fig.~\ref{fig:ladder}{\bf a}. We then have
\begin{equation}
  {\cal P}_\mathrm{even}^\alpha(r) = \langle l_i^\alpha\exp(i\pi\sum_{l=i+1}^{i+r-1}\tau_l^\alpha)l_{i+r}^\alpha\rangle,
  \label{eq:peven}
\end{equation} 
with $\tau_i^\alpha\mathord{=}S^\alpha_{i,1}+S^\alpha_{i+1,2}$ as above. Results for ${\cal P}_\mathrm{even}^z$ versus $\phi$ obtained from iDMRG
calculations with \HU\ 
are shown in Fig.~\ref{fig:AGY} as the light blue points. The \FK, \SPTA\ and \SPTB-phases are clearly defined by a non-zero 
${\cal P}_\mathrm{even}$. By applying $U_6^{-1}$ to the definition of ${\cal P}_\mathrm{even}$ it is straightforward to perform
similar calculations using \HKG\ by evaluating $U_6^{-1}({\cal P}_\mathrm{even}^\alpha(r))$. Even though the definition
${\cal P}_\mathrm{even}$ is heuristic, we interpret this result as a verification of the SPT nature of the \FK, \SPTA- and \SPTB-phases.

For the \FK-phase, it is also instructive to consider an even simpler heuristic string order correlation function, $\mathcal{Z}$ defined
as follows~\cite{feng2007characterization}:
\begin{equation}
  \mathcal{Z} = \langle\prod_i^{i+r}\sigma_i^z\rangle
  \label{eq:Z}
\end{equation} 
We consider this string correlator for \HU\ or equivalently to \HKG\ through application of $U_6^{-1}$. Results for $\mathcal{Z}$
are shown in Fig.~\ref{fig:AGY} versus $\phi$ (green points) as obtained from iDMRG calculations. Throughout the \FK-phase $\mathcal{Z}$ is almost identical to 1, dropping to zero at the transition to the \AG-phase. However, we note that $\mathcal{Z}$ remain sizable throughout much
of the \RSU-phase, reflecting its heuristic nature.

\subsection*{Edge states and response to magnetic field}\label{sec:edgestates}

Another signature of SPT phases is the presence of edge states under OBC related to 
a ground state degeneracy.
For the SPT phases in the K$\Gamma$ ladder, 
it is clear from the degeneracy of the entanglement spectrum that we need to consider different shapes of clusters (regular vs. slanted OBCs) for the different
SPT phases. In this section, we exclusively study the original Hamiltonian \HKG\ and do not consider \HU\ nor $H_{\mathrm{KQ}}$.
For \AG\ phase, we use $N\mathord{=}4n$ with the regular OBC
and for the remaining SPT phases, we use the slanted OBC with
$N\mathord{=}4n+2$ in order to have an equal number of the different bond types.

We first demonstrate the presence of edge states in the SPT phases. 
For the \AG\ phase, results for the 16 lowest states with the regular OBC 
at $\phi\mathord{=}0.85\pi$ are obtained using ED (see Supplementary Figure 7).
Four low-lying states below the gap are clearly present. With increasing $N$, these 4 states quickly become degenerate while
the gap stabilizes at a finite value. Similar results can be obtained for the \AK\ and \FK\ phase using the slanted cluster,
and a degeneracy of 4 is also observed for this case (see Supplementary Figure 8).  For the \SPTA- and \SPTB-phases, it has not been possible to produce reliable results in the same manner, likely due to the significantly larger correlation lengths.

\begin{figure}[t]
        \includegraphics[width=\columnwidth]{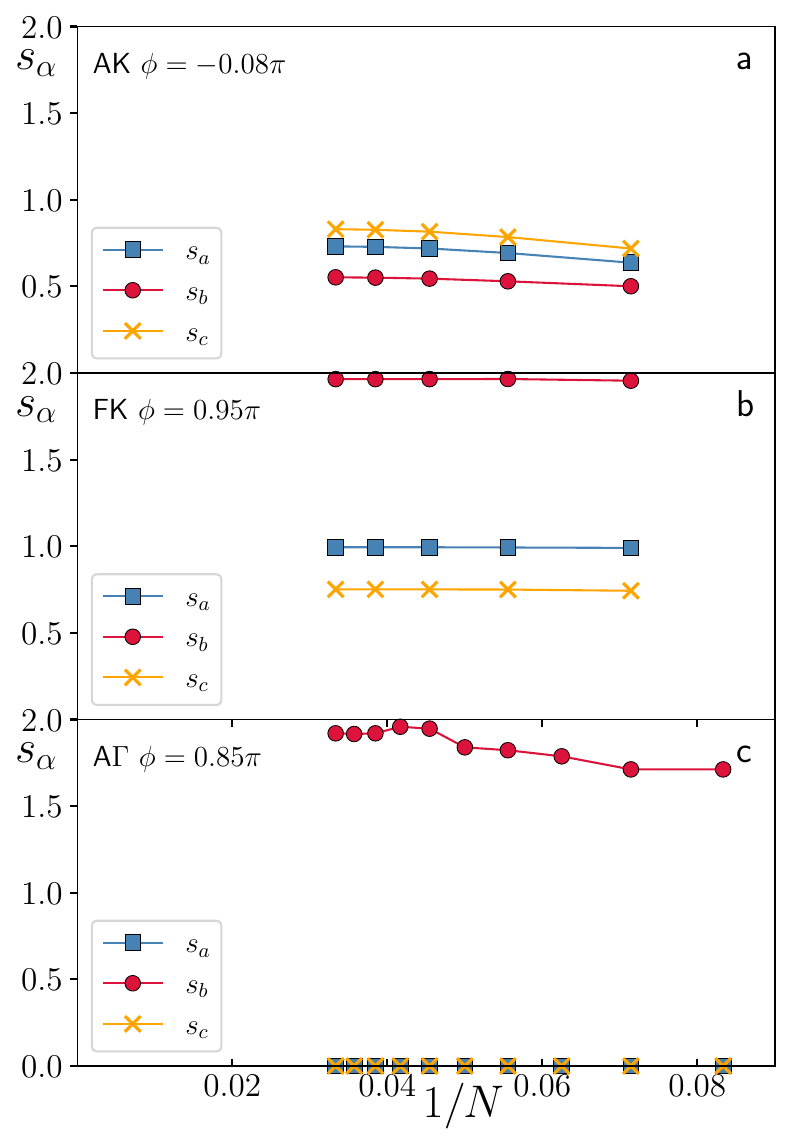}
  \caption{{\bf Eigenvalues of total spin in the \AK, \FK\ and \AG\ phases.} The eigenvalues $s_\alpha$=$s_{a,b,c}$ of the total spin $S_\alpha$=$\sum_i S^\alpha_i$ in the four degenerate ground-states, in zero field. Results are from ED.  
  {\bf a} \AK-phase at $\phi=-0.08\pi$ with $N\mathord{=}14,18\ldots 30$
  {\bf b} \FK-phase at $\phi=0.95\pi$ with $N\mathord{=}14,18\ldots 30$, {\bf c} \AG-phase at $\phi\mathord{=}0.85\pi$ with $N\mathord{=}12,14,\ldots 30$ (filled symbols).
  }
  \label{fig:EdgeSAGFK}
\end{figure}

To understand the nature of the edge-states, let us explore how
 the \AK,  \FK, and \AG\ phases respond to an external magnetic field. 
An external magnetic field introduces an additional term in the Hamiltonian of the form $H'\mathord{=}g_L\mu_B {\bf B}\cdot {\bf S}_\mathrm{tot}$, where ${\bf S}_\mathrm{tot}\mathord{=}\sum_i\mathbf{S}_i$, $g_L$ is the Land\'e factor and $\mu_B$ the Bohr magneton. 
Following Ref.~\cite{Liu2011} we denote the 4 states $\psi_1, \psi_2, \psi_3 $
and $\psi_4$ and consider $\mathbf{S}_{\mathrm{tot},\alpha}$ in this four-fold degenerate space by defining
\begin{equation}
    S^{\gamma\beta}_{\mathrm{tot},\alpha}=\langle \psi_\gamma|S_{\mathrm{tot},\alpha}|\psi_\beta\rangle,\ \ \gamma,\beta=1,2,3,4.
\end{equation}
Here, the components of the total spin $S_{\mathrm{tot},\alpha}$ are usually taken to be identical to $x,y,z$ but given the underlying honeycomb structure we shall find it useful to instead consider $\alpha$=$\mathbf{\hat{a}}$, $\mathbf{\hat{b}}$, $\mathbf{\hat{c}}$ corresponding to the three directions $[$11-2$]$, $[$1-10$]$ and
$[111]$ that correspond to the perpendicular and parallel to the z-bond, and perpendicular to the plane of the honeycomb (or n-leg brick-wall), respectively. One then finds that the eigenvalues of the matrices $(S^{\gamma\beta}_{\mathrm{tot},\alpha})$ for four corresponding degenerate states are simply given by
$(s_\alpha,-s_\alpha,0,0)$. 

Our ED results for $s_\alpha$ are shown in Fig.~\ref{fig:EdgeSAGFK}. For the \AK\ phase at $\phi\mathord{=-}0.08\pi$, shown in Fig.~\ref{fig:EdgeSAGFK}(a), we find $s_a\mathord{\sim}3/4$, $s_b\mathord{\sim}1/2$ 
and $s_c\mathord{\simeq}0.85$.
Similarly, for the \FK-phase  at $\phi\mathord{=}0.95\pi$, shown in Fig.~\ref{fig:EdgeSAGFK}(b), we find $s_a$=1, $s_b\mathord{\sim}2$ 
and $s_c$=3/4.  In both cases, we expect some variation in the values of $s_\alpha$ as $\phi$ is tuned.
We note that for both the \FK\ and \AG\ phase, the values of $s_\alpha$ quickly saturate at a small, finite value as $N$ is increased.
This is indicative of excitations localized at the edges as opposed to an actual magnetically ordered ground-state which should
show $s_\alpha$ continually growing with $N$. A calculation of $s_\alpha$ for the \SPTA\ and \SPTB-phases do not yield clear
results for the range of $N$ available with ED, as discussed in Supplementary Note 6.
In a realistic experimental setting, the presence of impurities will always create finite open segments of ladders, with a resulting Curie-law behavior. The response to an applied magnetic field is in that case highly anisotropic and
the low temperature Curie-law response should show a strong directional dependence~\cite{Liu2011,Liu2012,Chen2015} with $\chi_a(T)$, $\chi_b(T)$ and $\chi_c(T)$ clearly distinguishable. 
The results for the \AG-phase, shown in Fig.~\ref{fig:EdgeSAGFK}(c),
are even more intriguing. They are not only more anisotropic, but only $s_b$ is non-zero, approaching a value close to 2 at $\phi\mathord{=}0.85\pi$. At a slightly different point in the \AG-phase with $\phi\mathord{=}0.8\pi$ we instead find $s_b\mathord{\simeq}4/3$ but again only $s_b$ is non-zero.
This implies that the phase does not respond  at all to a field applied along the $\mathbf{\hat{a}}$ and $\mathbf{\hat{c}}$ directions,
effectively $g_a, g_c\mathord{\simeq}0$. 

\begin{figure}
        \includegraphics[width=\columnwidth]{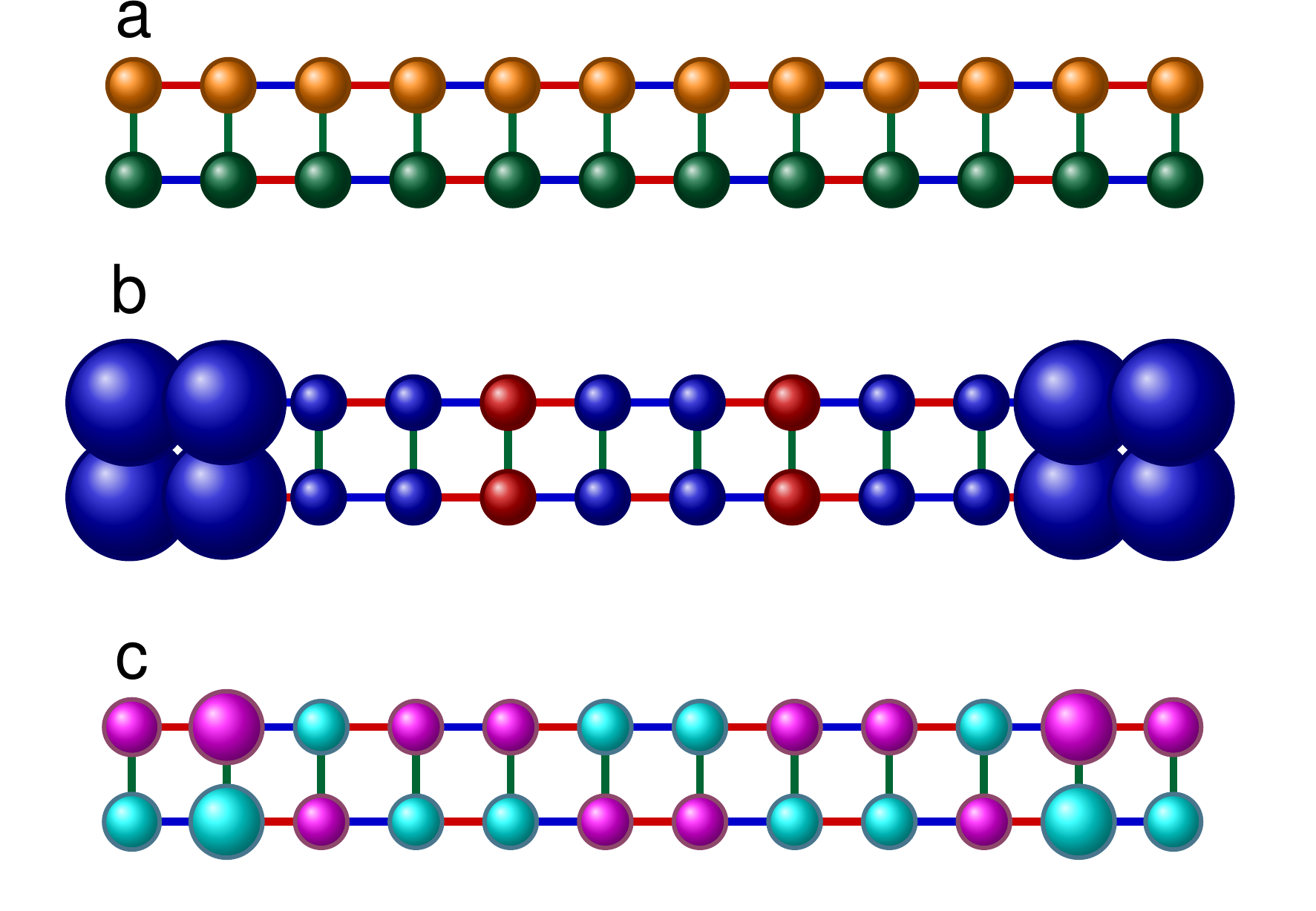}
  \caption{{\bf Ground state magnetization in an infinitesimal field in the \AG\ phase.} ED results with $N$=24 for $\langle S^{a,b,c}_i\rangle $ in the lowest state in the AG phase at $\phi/\pi$=0.85 with a field in the $\mathbf{\hat{b}}$-direction of strength 0.002. The 
  green, blue and cyan colors indicate positive values while orange, red, and pink indicate negative values. The size of the circles are proportional to the value of $\langle S^{a,b,c}_i\rangle$. {\bf a} $\langle S^a_i\rangle$, {\bf b} $\langle S^b_i\rangle$, {\bf c} $\langle S^c_i\rangle$.}
    \label{fig:Sabc}
\end{figure}

\begin{figure}[t]
        \includegraphics[width=\columnwidth]{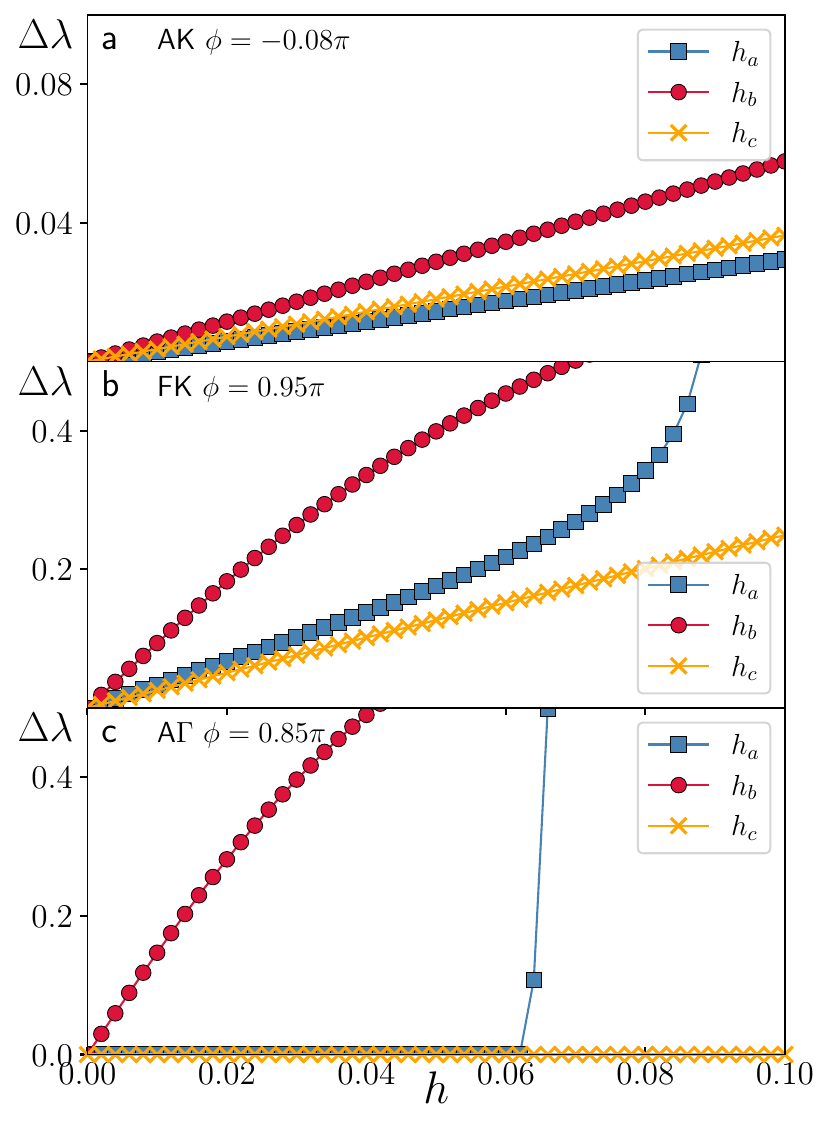}
  \caption{{\bf The Schmidt gap in the \AK, \FK\ and \AG\ phases}.
  The splitting of the entanglement spectrum $\Delta\mathord{=}\lambda_1\mathord{-}\lambda_2$ from iDMRG calculations with a field in the
  $\mathbf{\hat a}$, $\mathbf{\hat b}$ and $\mathbf{\hat c}$ directions.
  {\bf a} \AK-phase at $\phi=-0.08\pi$ using \RHOB.
  {\bf b} \FK-phase at $\phi=0.95\pi$ using \RHOB. 
  {\bf c} \AG-phase at $\phi\mathord{=}0.85\pi$ using \RHOA.
  }
  \label{fig:ESsplit}
\end{figure}

To gain a clearer picture of how the ladder in the \AG\ phase responds to a magnetic field applied along the $\mathbf{\hat{b}}$-direction, we have performed
ED calculations in the presence of a small field in the $\mathbf{\hat{b}}$-direction. The resulting site dependent magnetization $\langle S^{a,b,c}_i\rangle$ can
then easily be obtained for the $\mathbf{\hat{a}}, \mathbf{\hat{b}}$ and $\mathbf{\hat{c}}$-directions. Our results are shown in Fig.~\ref{fig:Sabc} for $N$=24 as obtained
from the ground-state with a small field in the $\mathbf{\hat{b}}$-direction of 0.002 at $\phi/\pi$=0.85 in the \AG-phase.
The green, blue and cyan colors represent positive expectation values, 
whereas orange, red and pink colors indicate negative values, with the size of the points proportional to the expectation value.
The response along the $\mathbf{\hat{a}}$, $\mathbf{\hat{c}}$ directions
is completely symmetric in the positive and negative directions, yielding $\sum_i S^{a,c}_i$=0. However, along the $\mathbf{\hat{b}}$-direction the response is much larger, and we clearly find $\sum_i S^{b}_i\neq$0
consistent with the results shown in the SI (See Fig. S9) (c).
With the field along the $\mathbf{\hat{b}}$-direction, sizeable excitations are visible at both ends of the ladder.

The response of the edge states to a magnetic field correlates with the lifting (or absence of lifting) of the degeneracy in the entanglement spectrum.
As previously discussed, non-trivial indices in the projective analysis can only arise if the degeneracy of all states in the entanglement spectrum (ES) is larger than one. This implies that if a finite strength of the perturbation is needed to remove the degeneracy, then a phase transition does not occur until that strength is reached and the phase persists till that point. On the other hand, if the degeneracy is lifted for any non-zero strength of the perturbation, the symmetry protection is broken without an associated phase transition.
We can then investigate the response of the degeneracy of the ES to a magnetic field in the $\mathbf{\hat a}$, $\mathbf{\hat b}$ and $\mathbf{\hat c}$ directions.
This is shown in Fig.~\ref{fig:ESsplit} for the \AK, \FK, and \AG\ phases. For simplicity, we focus exclusively on the difference in the two largest eigenvalues
$\Delta\mathord{=}\lambda_1\mathord{-}\lambda_2$, and in each case we employ the reduced density matrix that has a two-fold degeneracy at zero field. 

As can be seen in Fig.~\ref{fig:ESsplit}(a) the degeneracy in the ES for the \AK\ phase is immediately lifted by a field in any of the three directions which
correlates with the response of the edge-states (See Fig. S9(a) in the SI).
However, the response is rather weak, and a relatively large field has to be applied to see a significant splitting. For the \FK\ phase, we have a similar effect as shown in Fig.~\ref{fig:ESsplit}(b), but in this case the response is much stronger. However, for the \AG\ phase, where we show results in Fig.~\ref{fig:ESsplit}(c), it is clear that a field in the
$\mathbf{\hat a}$ direction
of around $h_a$=0.06 is needed to
lift the degeneracy of the ES. For a field applied in the $\mathbf{\hat c}$ direction, a significantly larger field is needed.  On the other
hand, a field in the $\mathbf{\hat b}$ direction immediately induces a large splitting in the entanglement spectrum even for infinitesimal field strengths.
Since the degeneracy remains intact in the $\mathbf{\hat a}$ and $\mathbf{\hat c}$ directions, we conclude that the SPT character of the \AG\ phase persists with respect to a field applied in the
$\mathbf{\hat a}$ and $\mathbf{\hat c}$ directions.

The \AG\ phase is then protected by the product of time-reversal ($\mathrm{TR}$) and $\pi$ rotation around the $\mathbf{\hat{b}}$-axis ($\mathcal{R}_{b}$), $\mathrm{TR}\times \mathcal{R}_{b}$, the only
remaining symmetry~\cite{Cen2023PRB} when the field is in the ${\mathbf{\hat a}}{\mathbf{\hat c}}$ plane, but broken when it is along the ${\hat b}$-axis.
Hence, if a field is applied in the ${\mathbf{\hat a}}{\mathbf{\hat c}}$ plane, a transition to the trivial polarized state can only occur at finite field strengths with potentially other phases 
intervening before the polarized state is encountered.
Several such transitions  were observed for the \AG\ phase (denoted by K$\Gamma$SL) for a field in the ${\mathbf{\hat c}}$ direction~\cite{Gordon2019,sorensen2021prx}.
On the other hand, the \FK\ phase is not protected by the $\mathrm{TR}\times \mathcal{R}_{b}$ symmetry, and if a field is applied in the ${\mathbf{\hat c}}$ direction the ES degeneracy is lost as shown in Fig.~\ref{fig:ESsplit}(b).
However, the field induced \FK\ phase can still be distinguished from the polarized state.

\section*{Discussion}
Our initial inquiry in this paper pertains to the nature of the \AG\ phase and whether there exists a defining quantity for its characterization.
For example, the Kitaev phases (\AK\ and \FK) in the ladder display the character of SPT phases.
It is likely that \AG\ is another SPT phase. If so, we expect all the signatures of the SPT such as the degeneracy of the entanglement spectrum, ground state degeneracy under OBC, and the presence of a SOP. 
Using the iDMRG, DMRG, and ED techniques, we indeed found that the entanglement spectrum is degenerate and there exists four-fold ground state degeneracy under the regular OBC in the \AG\ phase.
It is interesting to note that the same results were obtained for the Kitaev phases, \AK\ and \FK, but under the slanted OBC.

Despite such clear signatures of the SPT,
determining the corresponding SOP in the \AG\ phase has been challenging.
We found that the string order correlation function 
is related to ordinary local order in a regular correlation function in a model $H_{\mathrm{KQ}}$
obtained only after 
two consecutive unitary transformations. 
Hence, we term this order as 'twice' hidden.

To understand the symmetry that protects the \AG\ phase,
we also investigated the effects of the external magnetic field.
From the magnetic field response, we noted that the \AG\ phase is completely inert to the magnetic field when the field is applied in the ${\hat a}$ and ${\hat c}$ directions, which correspond to perpendicular to the z-bond and the ladder plane, respectively. 
Accordingly, the entanglement spectrum degeneracy remains intact when the field is applied in the ${\hat a}$ and ${\hat c}$ directions.
This is in contrast to the effect of applying the magnetic field along the ${\hat b}$ direction, i.e., parallel to the z-bond, which
immediately lifts the degeneracy of the entanglement spectrum.
We note that the product of $\mathrm{TR}$ and $\mathcal{R}_b$,  $\mathrm{TR}\times \mathcal{R}_{b}$ symmetry, is preserved
when the magnetic field is applied in the ${\mathbf{\hat a}}{\mathbf{\hat c}}$-plane, which is valid for the generic honeycomb Kitaev model beyond the ladder\cite{Cen2023PRB}.
Thus, we conclude that the \AG\ is protected by the $\mathrm{TR}\times \mathcal{R}_{b}$ symmetry.
Another intriguing implication from the magnetic field study is that the edge state in the \AG\ phase
is not a isotropic free spin-1/2 unlike the standard S=1 Haldane SPT.
They act like spinless modes under  the field in ${\hat a}$ and ${\hat c}$ directions. 
Further studies are needed to fully understand the nature of the zero-energy modes at the boundary of the system with the regular OBC.

 In the context of Kitaev materials, let us revisit our motivations for investigating the \AG\ phase in the ladder model.
As previously mentioned in the introduction, the majority of $d^5$ Kitaev materials prominently feature FM Kitaev and AFM $\Gamma$ interactions. However, ongoing debates persist regarding the specific phase that arises in this region. Several numerical studies have suggested the presence of magnetic disorder~\cite{Catuneanu2018npj,Gohlke2018prb,Gohlke2020PRR,Yamada2020prb,Lee2020Magnetic,
Luo2021npj,Gordon2019,Yilmaz2022PRR},  while others have indicated a magnetically ordered phase, such as a zig-zag order~\cite{Wang2019prl}.
Should a zig-zag order indeed be manifest in the 2D limit, we would anticipate observing the same ordering pattern in the 2-leg ladder, as the magnetic unit cell of the zig-zag can be captured in the ladder geometry. 
This is indeed the case for the Kitaev-Heisenberg ladder model, where the zig-zag, stripy, and FM ordered phases reported in the 2D honeycomb clusters are found in the ladder geometry~\cite{Catuneanu2018ladder}.

Our findings have substantiated the presence of disordered state in the \AG\ phase  of the 2-leg ladder, categorizing it as a SPT phase characterized by a SOP.
 The 2D limit can be constructed by stacking the ladders, and one possibility of the resulting 2D phase is a stacked SPTs with edge modes known as a weak-SPT~\cite{You2018PRB}.
However, the coupling between the ladders may generate a new phase or critical point.
It is interesting to note that
the evolution from a stacked weak-SPT chains to a gapless critical point supporting edge modes that do not hybridize with bulk modes was reported in the extended anisotropic Kitaev model approaching from the  dimer limit~\cite{Nanda2021PRB}.
Our findings hint at the possibility that as the 2D limit is approached, the \AG\ phase may become a 2D spin liquid, denoted as the K$\Gamma$ spin liquid.  
However, we cannot rule out a possibility of large unit cell~\cite{rayyan2021PRB} or incommensurate~\cite{Buessen2021} magnetic orders whose magnetic unit cells are beyond the ladder geometry, and a definitive resolution to this question remains a subject for future investigation.

\section*{Methods}
\subsection{Numerical Methods}
We use a fully parallelized implementation~\cite{LauchliED2011} of the Lanczos algorithm to perform the exact diagonalizations (ED)
of ladders with up to $N$=30 using both open and periodic boundary conditions.
In addition to exact diagonalizations, we use finite size density matrix renormalization group~\cite{White1992a,White1992b,White1993,Schollwock2005,Hallberg2006,Schollwock2011} (DMRG)
to study both the \KG\ model, Eq.~(\ref{eq:HKG}) and \HU\ under both periodic (PBC) and open (OBC) boundary conditions, with the main part of our results
obtained from the infinite DMRG~\cite{McCulloch2008,Schollwock2011} (iDMRG) variant of DMRG.
The iDMRG calculations were performed with unit cells of 12, 24 or 60 sites.
Typical precision for both DMRG and iDMRG are $\epsilon<10^{-11}$ with a bond dimension in excess of 1000.

\subsection{Energy susceptibility}
To determine the phase diagram we study the susceptibility derived from the ground-state energy per spin, $e_0$:
\begin{equation}
   \chi_{\phi}^e = -\frac{\partial^2 e_0}{\partial \phi^2}, \
\end{equation}
At a quantum critical point (QCP) it is known~\cite{Albuquerque2010} that,
for a finite system of size $N$, the energy susceptibility diverges as
\begin{equation}
  \chi^e \sim N^{2/\nu-(d+z)}.
\end{equation}
Here $\nu$ and $z$ are the correlation and dynamical critical exponents and $d$ is the dimension.
Hence,  $\chi^e$ only diverges at the phase transition if the critical exponent $\nu$ is smaller than $2/(d+z)$. 
For the present case we have $d$=$1$ and if we assume $z$=1, then a divergence is observed only if $\nu<1$.

\subsection{Entanglement Entropy and Spectrum, Schmidt gap}
When studying the ladder shown in Fig.~\ref{fig:ladder}{\bf a} it is important to realize that there are different ways of partitioning the system
in two partitions of size $x$ and $N-x$. 
This is crucial when considering the bipartite von Neumann entanglement entropy, $EE$, 
as well as for the entanglement spectrum~\cite{Li2008} of central importance for 
understanding topological properties~\cite{Li2008,Pollmann2010,Cho2017a,Cho2017b}. Both are obtained
from the spectrum of the reduced density matrix, $\rho_x$, of either one of the two partitions. Here we focus on two specific partitions
shown in Fig.~\ref{fig:ladder}{\bf a} as the red and blue dashed lines. 
With the numbering in Fig.~\ref{fig:ladder}{\bf a}, they correspond to either
an odd ($N/2$-1, red) or even ($N/2$, blue) number of sites in the partitions. 
We refer to the density matrix derived from the former case with $N/2$-1 as \RHOB\ and to the latter case with $N/2$ as \RHOA.
We mainly focus on the case where the number of
sites in the partition is close to the mid-point, either $N/2-1$ (\RHOB) or $N/2$ (\RHOA).
but when considering the bipartite entanglement entropy, we let the number in the partition
vary but only consider an even number of sites in the sub system corresponding to moving the blue dashed line in Fig.~\ref{fig:ladder}{\bf a} along the
ladder. For a subsystem, $A$, of size $x$ the entanglement entropy is defined by:
\begin{equation}
    EE(x) = -\mathrm{Tr}\rho_x\ln\rho_x.
\end{equation}
Our results for $EE(x)$ can be found in Supplementary Note 2.
The eigenvalues, $l_\alpha$, of the reduced density matrix, $\rho_x$, correspond to  the Schmidt decomposition, $l_\alpha$=$\lambda_\alpha^2$ and
thereby the entanglement spectrum~\cite{Li2008}, which then will depend on whether \RHOA\ or \RHOB\ from Fig.~\ref{fig:ladder}{\bf a} is used.
The Schmidt gap is then defined as $\lambda_1-\lambda_2$.

\section*{Data Availability}
The data that support the findings of this study are available at \doi{10.5281/zenodo.10443031}, or alternatively from the corresponding authors upon reasonable request.

\section*{Code Availability}
The code used to generate the data used in this study is available from the corresponding author upon reasonable request.

\begin{acknowledgments}
We thank A. Catuneanu and J. Gordon for fruitful discussions during the initial stages of this work.
We acknowledge the support of
the Natural Sciences and Engineering Research Council of Canada (NSERC) through Discovery
Grants (No. RGPIN-2017-05759 and No. RGPIN-2022-04601).  
H.-Y.K. acknowledges the support of CIFAR and the Canada Research Chairs Program. 
This research was enabled in part by support provided by SHARCNET (sharcnet.ca) and the Digital Research Alliance of Canada (alliancecan.ca).
Part of the numerical
calculations were performed using the ITensor library~\cite{itensor}.
\end{acknowledgments}

\section*{Author contributions}
Density-matrix renormalization group and exact diagonalization calculations were performed by E.S.S. 
H.-Y. K.  initiated the project.  All authors managed the project and wrote the manuscript. 

\section*{Competing Interests}
The authors declare no Competing Financial or Non-Financial Interests.

\ifarXiv
    

\section*{Supplementary material}

# Supplementary Information for:

# Twice Hidden String Order in the Spin-1/2 Kitaev-Gamma Ladder

Erik S. Sørensen,[1] and Hae-Young Kee[*] [2, 3]

[1]*Department of Physics, McMaster University,*

*Hamilton, Ontario L8S 4M1, Canada*

[2]*Department of Physics, University of Toronto, Ontario M5S 1A7, Canada*

[3]*Canadian Institute for Advanced Research,*

*CIFAR Program in Quantum Materials, Toronto, ON M5G 1M1, Canada*

(Dated: January 15, 2024)

**CONTENTS**

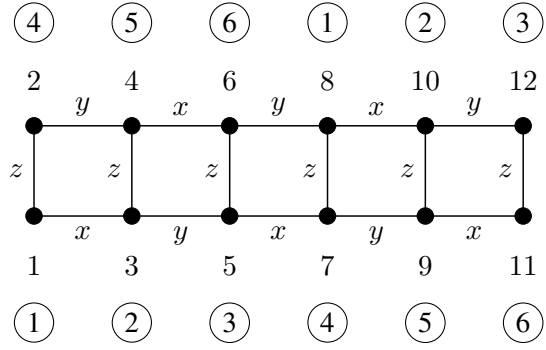

Supplementary Figure 1. The $U_6$ transformation applied to the Kitaev-$\Gamma$ ladder. The circled numbers indicate the sub-lattice.

## I. SUPPLEMENTARY NOTE 1: THE $U_6$ TRANSFORMATION APPLIED TO THE LADDER

In this supplementary note, we briefly summarize the $U_6$ transformation used in the main paper. Following Ref. [1], we define the $U_6$ transformation on 6 sub-lattices as follows:

① $(x, y, z) \to (x', y', z')$

② $(x, y, z) \to (-x', -z', -y')$

③ $(x, y, z) \to (y', z', x')$

④ $(x, y, z) \to (-y', -x', -z')$

⑤ $(x, y, z) \to (z', x', y')$

⑥ $(x, y, z) \to (-z', -y', -x')$

The sub-lattices can then be assigned to the ladder as shown in Supplementary Figure 1. This then leads to the form of the transformed Hamiltonian $H_{\text{K}\Gamma}^{U_6}$ shown in Supplementary Figure 5 with bond interactions defined in Eq. (3) of the main paper. Note that $U_6^2 \neq \mathbb{1}$. The inverse transformation $U_6^{-1}$ is then given by:

① $(x, y, z) \to (x', y', z')$

② $(x, y, z) \to (-x', -z', -y')$

③ $(x, y, z) \to (z', x', y')$

④ $(x, y, z) \rightarrow (-y', -x', -z')$

⑤ $(x, y, z) \rightarrow (y', z', x')$

⑥ $(x, y, z) \rightarrow (-z', -y', -x')$

## II. SUPPLEMENTARY NOTE 2: GAPS AND CORRELATION LENGTH

In this supplementary note, we discuss the spin gap and associated correlation length in the SPT$_\alpha$, SPT$_\beta$, A$\Gamma$ and F$K$ phases. Here, we show that all 4 phases are gapped with a finite correlation length, although the correlation length int the SPT$_\beta$ and parts of the A$\Gamma$ phase is very large.

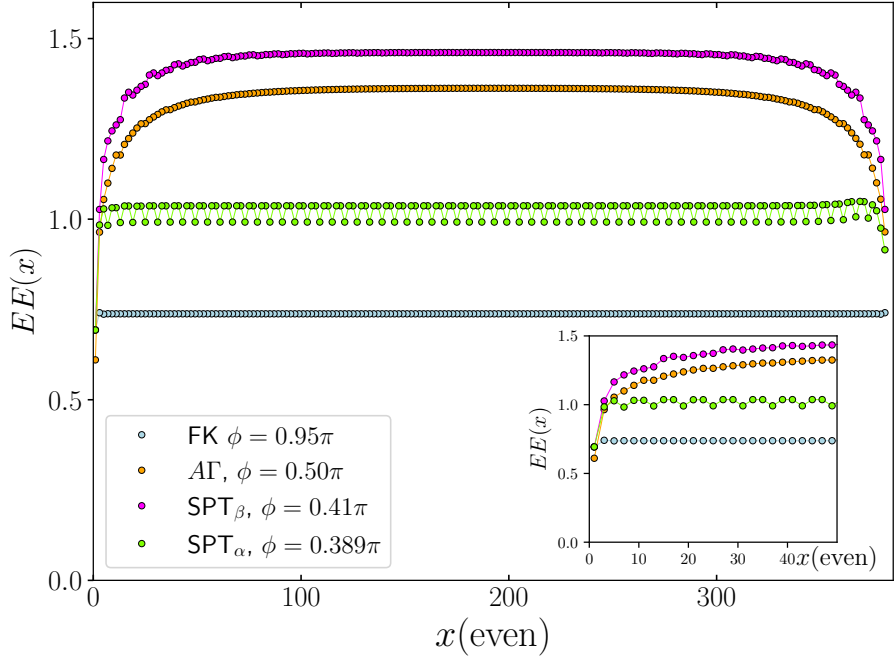

Supplementary Figure 2. The bipartite entanglement entropy versus x for a range of values of $\phi$. Results are from finite size DMRG calculations with N=384. Results are shown only for even $x$, corresponding to partitioning the chain similarly to the dashed blue line in Fig. 1**a** of the main paper. The flatness of $EE(x)$ around $x=N/2$ is consistent with the presence of a gap. The inset shows a close up of the results at the left edge.

In the most common scenario, we expect phases to be gapped. This is known to be the case for the A$\Gamma$ and F$K$ phases, where a clear gap is present in DMRG calculations with PBC [2, 3].

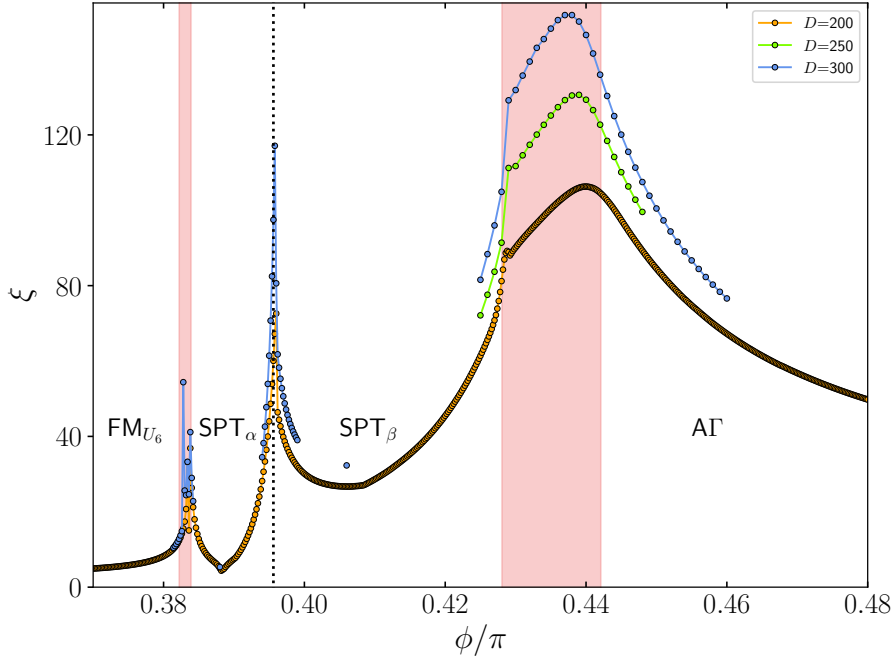

Supplementary Figure 3. The correlation length $\xi$ versus $\phi/\pi$ as determined from the transfer matrix in iDMRG with a bond dimension of $D=200, 250$ and $300$.

Similar calculations for the $SPT_\alpha$ and $SPT_\beta$ phases do not yield a clear result. We have therefore studied the bipartite entanglement entropy $EE(x)$ for a sub partition $A$ of linear extent $x$. It is known that for a gapless, one-dimensional conformal model $EE(x)$ in the limit $N \to \infty$ should diverge logarithmically [4, 5]

$$EE(x) \sim c \ln x, \qquad (1)$$

with $c$ the central charge. However, for a system away from criticality $EE(x)$ should instead be roughly constant, behaving as

$$EE(x) \sim c \ln \xi, \qquad (2)$$

where $\xi$ is the correlation length. It follows that $EE(x)$ plotted versus $x$ should approach a constant value independent of $x$ for $x \sim N/2$ if a gap is present. Our results for $EE(x)$ are shown in Supplementary Figure 2 and the gap present in the $A\Gamma$ and $FK$ phases are immediately visible since the corresponding results for $EE(x)$ are almost independent of $x$. Furthermore, $EE(x)$ for $FK$-phase saturates at a relatively small value, indicating a sizable gap. The results for the $A\Gamma$-phase saturate at a higher value, but it should be noted that the results are obtained at the point $\Gamma=1$ where the gap is relatively small. Elsewhere in the $A\Gamma$-phase, the saturation value is comparatively smaller. The more interesting results are for the $SPT_\alpha$ and $SPT_\beta$ phases, which also

clearly saturate, indicating the presence of a gap. In the case of the SPT$_\alpha$-phase, the saturation value is relatively small, hinting at a discernible gap larger than for the SPT$_\beta$-phase.

For quasi one-dimensional systems, another measure of the gap can be obtained from the correlation length, which should be inversely related to the gap. The correlation length, $\xi$, can be obtained from the second eigenvalue of the transfer matrix in iDMRG calculations using matrix product states (MPS). As a function of the bond dimension, $D$, the size of the transfer matrix is $D^2$, limiting the range of $D$ that can be used for determining $\xi$. Our results for $\xi$ are shown in Supplementary Figure 3 for the SPT$_\alpha$, SPT$_\beta$ and A$\Gamma$-phases. Results are shown for $D=200, 250$ and 300. While the correlation length in the SPT$_\beta$-phase is sizeable, the correlation length in the SPT$_\alpha$-phase is much smaller, consistent with the presence of a well-defined gap in both phases. However, it is also clear from the results in Supplementary Figure 3 that the correlation length is also rather small in the FM$_{U_6}$-phase, a fact we turn to next.

### III. SUPPLEMENTARY NOTE 3: SPIN GAP IN THE FM$_{U6}$ PHASE

In this note, we discuss the spin gap in the FM$_{U_6}$ phase. At the special point $\phi=\pi/4$, the $U_6$ transformation discussed in supplementary note 1 maps the ladder to the *ferro*magnetic Heisenberg ladder. Hence, the name FM$_{U_6}$. At this particular point, we therefore know that the model does not have a gap. However, as we show in this note, the point $\phi=\pi/4$ is a unique point in the phase and symmetry breaking terms that arise as soon as $\phi \neq \pi/4$ immediately induce a gap and a finite correlation length.

In Supplementary Figure 4 we show results for the spin gap in the FM$_{U_6}$ phase at $\phi=0.35\pi$. The smaller colored points are obtained from exact diagonalization (ED) results with open boundary conditions (OBC) for $N = 10, 12, 14, 16, 18 \ldots 30$ with a regular shaped cluster. Results are shown for the first 16 levels relative to the ground-state energy. The coloring is only meant to facilitate a way of distinguishing the points, as are the lines connecting the points. Clearly, the results for $N = 12, 18, 24$ and 30, where $N$ is a multiple of 6, are distinct from the other values of $N$. In this case, when $N = 6n$ there is a clear gap present above a ground-state that is two-fold degenerate. We have confirmed this by performing DMRG calculations with OBC out to $N = 96$. For $N < 30$ the DMRG and ED results are in complete agreement, and for $N > 30$ the DMRG show that the gap converges to a finite value slightly above 0.04. The results are consistent with a two-fold degenerate magnetically ordered ground-state and a spin gap. As already mentioned, the

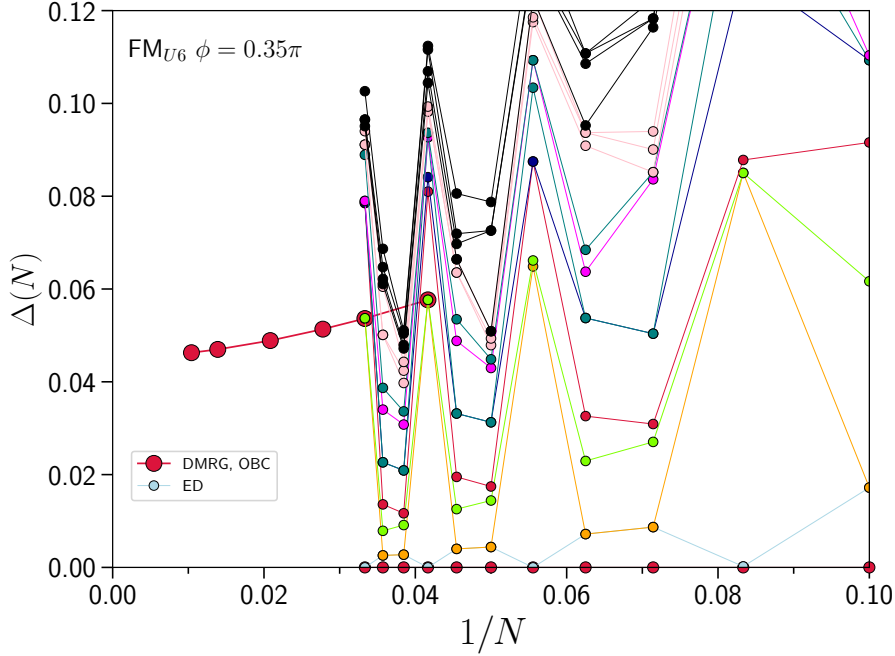

Supplementary Figure 4. ED results for the 16 lowest states with OBC in the FM$_{U6}$ phase at $\phi = 0.35\pi$ for $N = 10, 12, 14, 16, 18 \ldots 30$ shown alongside DMRG results with OBC (large red circles) and $N = 24, 30, 46, 48, 72, 96$. The lines and the coloring of the ED results only serve as a guide to the eye.

presence of a finite correlation length for a quasi one-dimensional system, implies the presence of a gap. As can be seen in Supplementary Figure 3, the correlation length in the FM$_{U_6}$-phase is rather small, indicating a finite gap. Additional calculations show that the correlation length only diverges at the point $\phi=\pi/4$ where $K=\Gamma$ and the mapping to the ferromagnetic ladder is exact. We have confirmed the presence of the spin gap in the FM$_{U_6}$-phase for $K\neq\Gamma$, by performing both ED and finite size DMRG calculations with both OBC and PBC. The use of PBC is of relatively limited use since, as can be seen from the $U_6$ transformation, one needs to consider systems with mod($N$,12)=0 where the ground-state is 6-fold degenerate. However, from the above analysis it is clear that with OBC the situation is more tractable since one only needs to consider systems with mod($N$,6)=0 where the ground-state only has a degeneracy of 2. Extrapolating the DMRG results for OBC to the thermodynamic limit can easily be done, and we can then obtain the spin gap throughout the FM$_{U_6}$-phase. This is shown in Supplementary Figure 5, where a modest spin gap can be seen developing as soon as $K\neq\Gamma$, going to zero at the point $\phi=\pi/4$.

6

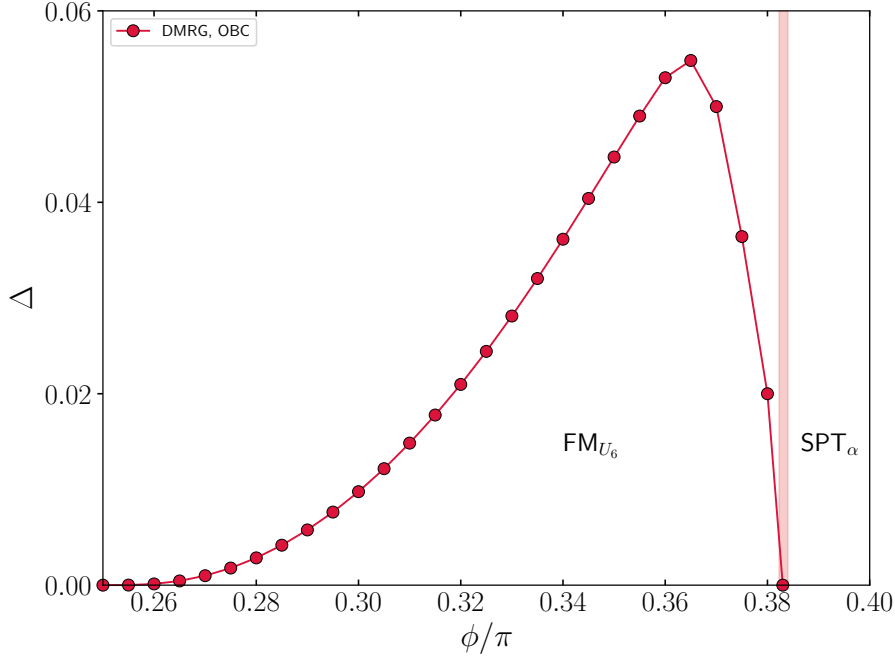

Supplementary Figure 5. The spin gap in the $FM_{U_6}$-phase as obtained from extrapolations to the thermodynamic limit of DMRG calculations with OBC and $N$ a multiple of 6, $N=12, 18, 24, \ldots$. Results are shown only for the part of the $FM_{U_6}$-phase where $\phi > \pi/4$. We expect results to be roughly identical for $\phi < \pi/4$.

## IV. SUPPLEMENTARY NOTE 4: MAPPING TO THE KQ MODEL

In this note, we discuss in more detail the mapping from $H_{K\Gamma}^{U_6}$ to $H_{KQ}$ through the application of the unitary operator $W$. Following Ref. [3], we apply the unitary operator $W$ from Eq. (14) (from the main paper) to the $U_6$ transformed Hamiltonian, $H_{K\Gamma}^{U_6}$ with regular OBC. We see that on the vertical bonds of the ladder $W$ leaves all interactions unchanged. For the remaining bonds, we

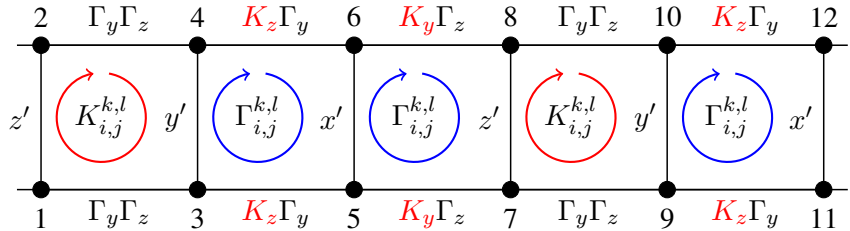

Supplementary Figure 6. $H_{KQ}$: The $K\Gamma$ ladder after the $U_6$ transformation followed by the $W$ transformation

7

introduce additional notation for how $W$ transforms the bonds:

$$\begin{aligned}
K_z\Gamma_y &: KS_i^zS_j^z + \Gamma S_i^yS_j^y \\
K_y\Gamma_z &: KS_i^yS_j^y + \Gamma S_i^zS_j^z \\
\Gamma_y\Gamma_z &: \Gamma S_i^yS_j^y + \Gamma S_i^zS_j^z \\
\Gamma_{i,j}^{k,l} &: 4\Gamma(S_k^zS_l^zS_i^yS_j^y + S_k^yS_l^yS_i^zS_j^z) \\
K_{i,j}^{k,l} &: 4K(S_k^zS_l^zS_i^yS_j^y + S_k^yS_l^yS_i^zS_j^z)
\end{aligned} \quad (3)$$

With this notation in hand, we can summarize the effect of the $W$ transformation on the $U_6$ transformed ladder in the manner shown in Supplementary Figure 6. The resulting Hamiltonian, denoted $H_{KQ}$, can be drawn in the manner shown in Supplementary Figure 6 where a relabelling of some sites has been done for clarity to undo a 'twisting' of the legs induced by the $W$ operator.

## V. SUPPLEMENTARY NOTE 5: SPIN GAP AND GROUND-STATE DEGENERACY IN THE A$\Gamma$ AND F$K$ PHASES

The most common scenario for SPT phases is the presence of a spin gap above a degenerate ground-state when open boundary conditions are applied. In this note we show that this scenario applies to the A$\Gamma$ and F$K$ phases, in both cases finding a four-fold degenerate ground-state under open boundary conditions and a well-defined spin-gap. However, to observe this degeneracy, *slanted* boundary conditions have to be applied for the F$K$ phase, corresponding to Fig. 4**b** in the main paper. In Supplementary Figure 7 we show ED results for the spin gap to the first 16 states in the A$\Gamma$-phase at $\phi = 0.85\pi$ with $N$=12, 14, 16, 18 ... 30 obtained using a regular shaped cluster, The points shown as large red circles are DMRG results with OBC and $N$=24, 30, 46, 48, 72, 96. The agreement between ED and DMRG results at small $N$ is almost exact, and a clear gap is visible above a 4-fold degenerate ground state.

In Supplementary Figure 8 we show results for the spin gap in the F$K$-phase at $\phi = 0.95\pi$ with OBC for $N$=14,18,22,26,30 using a *slanted* cluster. The colored points are from exact diagonalization with OBC for $N$=14,18,22,26,30 using a *slanted* cluster. Results are shown for the first 16 levels relative to the ground-state energy. The coloring of the points and the lines connecting them are only meant as an aide and are not related to physical quantum numbers. In this case, even for the smaller system sizes, is a 4-fold degenerate ground-state observed. Due to the almost complete degeneracy, it is not possible to discern the 4 separate points at zero energy

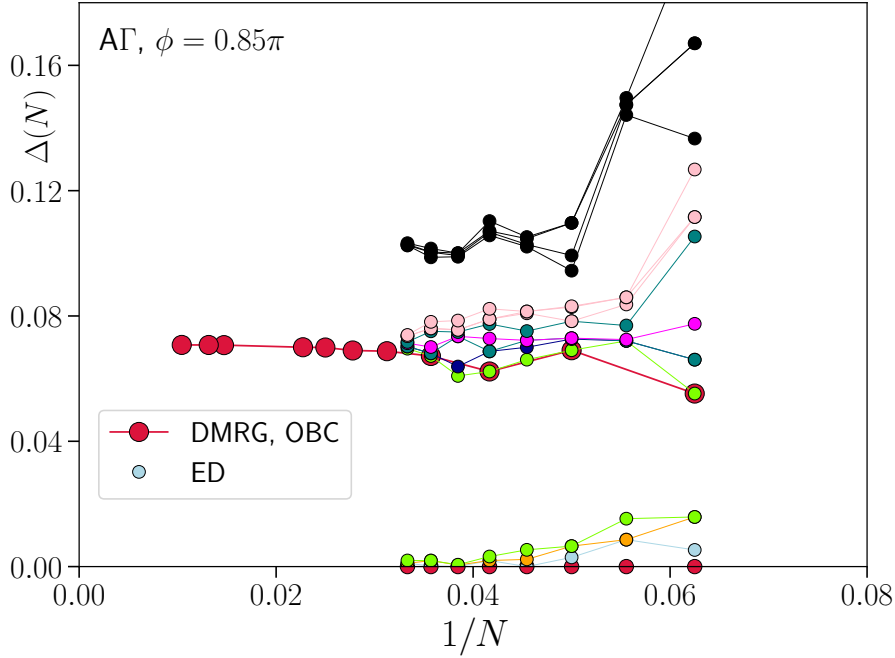

Supplementary Figure 7. ED results for the 16 lowest states with OBC in the AΓ phase at $\phi=0.85\pi$ for $N=12, 14, 16, 18 \ldots 30$ shown alongside DMRG results with OBC (large red circles) and $N=24, 30, 46, 48, 72, 96$. The lines and the coloring of the ED results only serve as a guide to the eye. A regular shaped cluster is used and energies are relative to the ground state energy.

in Supplementary Figure 8. The larger red circles are from DMRG with OBC for clusters out to $N=102$. The gap quickly converges to $\Delta = 0.119687771$.

## VI. SUPPLEMENTARY NOTE 6: MAGNETIZATION IN THE GROUND-STATE SUBSPACE OF THE $SPT_\alpha$ AND $SPT_\beta$ PHASES

In this supplementary note, we discuss what happens when the total magnetization is studied in the 4-fold degenerate sub-space of the ground-states by diagonalizing the total magnetization in the corresponding sub-space for the $SPT_\alpha$ and $SPT_\beta$ phases. We have determined the 4 lowest lying states using ED in the $SPT_\alpha$ and $SPT_\beta$ phases. The four states are only degenerate in the thermodynamic limit and, as one might expect, they do in general not show any kind of on-site ordering. As described in the main text, if a perturbation in the form of an external magnetic field is introduced, we need to consider an additional term in the Hamiltonian of the form $H'=g_L\mu_B \mathbf{B} \cdot \mathbf{S}_{tot}$, where $\mathbf{S}_{tot}=\sum_i \mathbf{S}_i$, $g_L$ is the Landé factor and $\mu_B$ the Bohr magneton. Following Ref. [6]

9

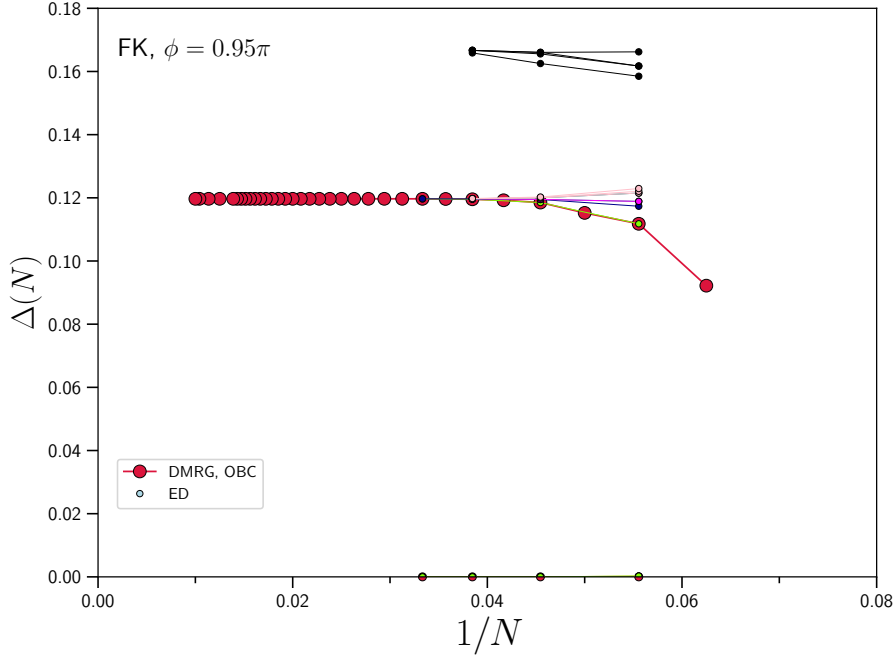

Supplementary Figure 8. ED results for the lowest states with OBC in the FK phase at $\phi = 0.95\pi$ for $N = 14, 18, 22, 26, 30$ shown alongside DMRG results with OBC (large red circles) and $N = 18, 22, 26, 30, \ldots 102$. The lines and the coloring of the ED results only serve as a guide to the eye. A slanted cluster is used and energies are relative to the ground state energy.

we denote the 4 states $\psi_1, \psi_2, \psi_3$ and $\psi_4$ and consider $\mathbf{S}_{\text{tot},\alpha}$ in this four-fold degenerate space by defining

$$S^{\gamma\beta}_{\text{tot},\alpha} = \langle \psi_\gamma | S_{\text{tot},\alpha} | \psi_\beta \rangle, \quad \gamma, \beta = 1, 2, 3, 4. \tag{4}$$

Here, the components of the total spin $S_{\text{tot},\alpha}$ are usually taken to be identical to $x, y, z$ but given the underlying honeycomb structure we shall find it useful to instead consider $\alpha = \hat{\mathbf{a}}, \hat{\mathbf{b}}, \hat{\mathbf{c}}$ corresponding to the three directions [11-2], [1-10] and [111]. One then finds that the eigenvalues of the matrices $(S^{\gamma\beta}_{\text{tot},\alpha})$ are simply given by $(s_\alpha, -s_\alpha, 0, 0)$. For, the $S=1$ Haldane chain, one can determine $s_x = s_y = s_z = 1$ and a small magnetic field applied along any direction will therefore split the ground-state degeneracy. The finding of $s_\alpha = 1$ for any direction is in this case taken to be consistent with the presence of two $S=1/2$ excitations at each end of the $S=1$ chain.

In the main paper, the eigenvalues $s_\alpha = s_{a,b,c}$ were discussed for the A$K$, A$\Gamma$ and F$K$ phases. We have repeated the calculations of the eigenvalues $s_\alpha = s_{a,b,c}$ for the SPT$_\alpha$ and SPT$_\beta$ phases. The results are shown in Supplementary Figure 9. In this case, the significantly larger correlation

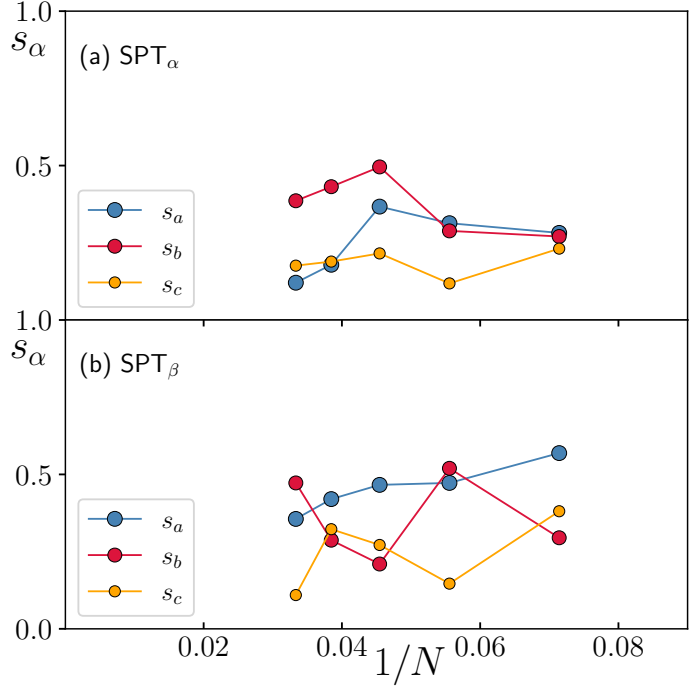

Supplementary Figure 9. The eigenvalues $s_\alpha = s_{a,b,c}$ of the total spin $S_\alpha = \sum_i S_i^\alpha$ in the four degenerate ground-states. Results are from ED with OBC in zero field. (a) SPT$_\alpha$-phase at $\phi = 0.388\pi$ with $N=14, 18 \ldots 30$. (b) SPT$_\beta$-phase at $\phi = 0.41\pi$ with $N=14, 18 \ldots 30$.

lengths hinder a clear interpretation of the results. Even though we find non-zero values for all $s_\alpha$ the trend with system size $N$ is not clear, nor systematic, and we cannot determine if any of them remain non-zero in the thermodynamic limit or attain a finite value. In both cases, we also expect some variation in the values of $s_\alpha$ as $\phi$ is tuned, further complicating the analysis.


\fi

\end{document}